\documentclass[number,sort&compress,times,review]{elsarticle}

\usepackage{graphicx,times,amsmath} 

\usepackage{textcomp} 

\usepackage{amsthm}
\usepackage{amssymb}
\usepackage{algorithm} 
\usepackage{xcolor}
\usepackage{mathrsfs}
\usepackage{url}
\usepackage{amsmath} 
\usepackage{subfigure}
\usepackage{caption}
\usepackage{float}
\usepackage{algorithmic}
\usepackage{multirow}
\usepackage{graphics}
\usepackage{hyperref}

\usepackage{epstopdf}

\usepackage{tabularx}

\theoremstyle{definition}

\theoremstyle{plain}

\usepackage{amsthm}
\usepackage{graphicx}
\usepackage{amssymb}
\usepackage{amsmath}
\usepackage{algorithm} 
\usepackage{algorithmic} 
\usepackage{multirow}
\usepackage{xcolor}
\usepackage{graphics}
\usepackage{mathrsfs}
\usepackage{hyperref}

\usepackage{subfigure}
\usepackage{float}
\usepackage{bm}
\usepackage{epstopdf}

\usepackage{tabularx}

\usepackage{balance}

\usepackage{tablefootnote}
\usepackage{threeparttable}

\usepackage{footnote}

\usepackage{makecell}

\usepackage{tablefootnote}

\theoremstyle{definition}

\newcommand{\highlight}[1]{\textcolor{black}{#1}}
\newcommand{\highlightR}[1]{\textcolor{black}{#1}}

\journal{Information Sciences}

\begin{document}
	
	\begin{frontmatter}
		
		

			\title{An Integrated \textit{Optimization + Learning} Approach to Optimal Dynamic Pricing for the Retailer with Multi-type Customers in Smart Grids\tnoteref{label-title}}
		
		

		 \author[label1,label4]{Fanlin Meng\corref{cor1}}

		 \author[label2]{ Xiao-Jun Zeng} 
		 	
		 \author[label3]{Yan Zhang}
		  
		 \author[label4]{Chris J. Dent}
		 
		  \author[label5] {Dunwei Gong}
		  
		  \tnotetext[label-title]{\textcopyright 2018. This manuscript version is made available under the CC-BY-NC-ND 4.0 license \\http://creativecommons.org/licenses/by-nc-nd/4.0/. \; Please cite this accepted article as: Fanlin Meng, Xiao-Jun Zeng, Yan Zhang, Chris J. Dent, Dunwei Gong, An Integrated Optimization + Learning Approach to Optimal Dynamic Pricing for the Retailer with Multi-type Customers in Smart Grids, Information Sciences (2018), doi: 10.1016/j.ins.2018.03.039}
		  
		  \cortext[cor1]{Corresponding author. Tel.: +44 131 650 5069; Email: Fanlin.Meng@ed.ac.uk.}
		  
		 \address[label1]{School of Engineering and Computing Sciences, Durham University, Durham DH1 3LE, UK}
		 
		 \address[label2]{School of Computer Science, The University of Manchester, Manchester M13 9PL, UK}
		 
		 \address[label3]{ College of Information System and Management, National University of Defense Technology, Changsha 410073, China}
		 
		 \address[label4]{School of Mathematics, University of Edinburgh, Edinburgh EH9 3FD, UK}
		 
		 \address[label5]{School of Information and Control Engineering, China University of Mining and Technology, Xuzhou 221116, China}

		\begin{abstract}
				In this paper, we consider a realistic and meaningful scenario in the context of smart grids where an electricity retailer serves three different types of customers, i.e., customers with an optimal home energy management system embedded in their smart meters (C-HEMS), customers with only smart meters (C-SM), and customers without smart meters (C-NONE).  The main objective of this paper is to support the retailer to make optimal day-ahead dynamic pricing decisions in such a mixed customer pool.  To this end, we propose a two-level decision-making framework where the retailer acting as upper-level agent firstly announces its electricity prices of next 24 hours and customers acting as lower-level agents subsequently schedule their energy usages accordingly. For the lower level problem, we model the price responsiveness of different customers according to their unique characteristics. For the upper level problem, we optimize the dynamic prices for the retailer to maximize its profit subject to realistic market constraints. The above two-level model is tackled by genetic algorithms (GA) based distributed optimization methods while its feasibility and effectiveness are confirmed via simulation results.  
		\end{abstract}

		
	\end{frontmatter}

	\section{Introduction} \label{Section-intro}

	With the large-scale deployment of smart meters and two-way communication infrastructures, dynamic pricing based demand response and \highlight{demand-side management programs} \cite{siano2014demand} \cite{esther2016survey} have attracted enormous attentions from both academia and industry and are expected to bring great benefits to the whole power system. \highlight{For dynamic pricing, price changes between different time segments. Real-time pricing (RTP), time-of-use  pricing (ToU) and critical-peak pricing (CPP) are commonly used dynamic pricing strategies \cite{khan2016load}. There are emerging studies on how to design optimal dynamic pricing strategies within the last decade. For instance, a residential implementation of CPP was investigated in \cite{herter2007residential} and an optimal ToU pricing for residential load control was proposed in \cite{datchanamoorthy2011optimal}. An optimal RTP algorithm based on utility maximization for smart grid was proposed in \cite{samadi2010optimal}. More recently, a game theory based dynamic pricing method for \highlight{demand-side management} was proposed in \cite{srinivasan2017game} where different pricing strategies such as RTP and ToU are evaluated for the residential and commercial sectors.} 
	
	Many existing studies on dynamic pricing based demand response and \highlight{demand-side management} assume that customers are installed with home energy management systems (HEMS) in their smart meters, i.e. an optimization software which is able to help customers maximize their benefits such as minimizing their payment bills. For instance, references \cite{mohsenian2010optimal, mohsenian2010autonomous, paterakis2015optimal,adika2013autonomous, wu2014real,  zhang2016model} propose different HEMS for customers to help them deal with the dynamic pricing signals. Instead of focusing on the single level customer-side optimization problems, references \cite{zugno2013bilevel, meng2013stackelberg, qian2013demand,chai2014demand, wei2015energy,yu2016real,yu2016supply} deal with how a retailer determines retail electricity prices based on the expected responses of customers where they model the interactions between a retailer and its customers as a Stackelberg game or bilevel optimization problem where HEMS are assumed to have been installed for all customers.

	In contrast, \cite{hosking2013short} \cite{meng2016profit} investigate the electricity consumption behavior of customers who have installed smart meters without HEMS embedded (C-SM).  Even without HEMS installed, these customers can easily get access to price information and their electricity consumption data through smart meters, and are likely to respond to dynamic price signals. On the other hand, with smart meters and  two-way communication, the retailer is able to identify each customer's energy consumption patterns based on history smart meter data. For instance, \cite{hosking2013short} presents a stochastic regression based approach to predict the hourly electricity consumption of each residential customer from historical smart meter data in the context of dynamic pricing. \cite{meng2016profit} proposes two appliance-level electricity consumption behavior models for each customer in the context of dynamic pricing with the premise that appliance-level electricity consumption data at each hour can be disaggregated from household-level smart meter data using non-intrusive appliance load monitoring (NILM) \cite{hart1992nonintrusive}.

	In addition to C-HEMS and C-SM, it is also unavoidable that some customers do not have smart meters installed (C-NONE). Therefore, these customers do not have \textit{direct access} to electricity prices \footnote{Although customers of this type could not receive price signals directly due to the unavailability of smart meters, the electricity price information is usually open to the public through other  sources (e.g., their retailer's website) and customers of this type might take advantage of this.} and their history consumption, and  are more likely to have a relatively low demand elasticity. On the other hand, without smart meters, the retailer does not have accurate consumption data of each individual customer but only the aggregated demand data of all customers. As a result, an aggregated demand model is needed to forecast the total demand of C-NONE. Existing research on aggregated demand modelling in the context of dynamic pricing include artificial intelligence based approaches \cite{yun2008rbf,khotanzad2002neuro,ren2016random} and demand elasticity based approaches \cite{kirschen2000factoring, ma2015demand, gomez2012learning, thimmapuram2013consumers}. 
	
	Although the above and other unlisted studies have provided valuable insights on how to model customers' demand patterns in the context of dynamic pricing and smart grids, they all consider scenarios where only one single type of customers exist. \highlight{However, there will be situations when several types of customers with different level of demand elasticity (e.g., C-HEMS, C-SM, and C-NONE) coexist in the market, especially during the transition phase of smart metering and smart grids (e.g., the smart meter roll-out in UK is expected to finish by 2020 \cite{UKSM2017}). Considering customer segmentation and differences in pricing has been extensively studied in retail sectors such as broadband and mobile phone industry \cite{fu2017designing}, but has not received much attention in energy and power system sector mainly due to the flat price regulation in the electricity retail market and lack of demand flexibility among customers. Nonetheless, with the liberalization of retail market and the development of smart grids and demand side management, situations start changing recently. For instance, the importance of considering customer segments has been demonstrated in some recent studies \cite{flath2012cluster} \cite{asadinejad2016residential}. More specifically, the benefit of capturing different customer groups with distinctive energy consumption patterns to the electricity tariff design was firstly described in \cite{flath2012cluster}.  Further, \cite{asadinejad2016residential} proposed an approach to group customers into different clusters (types) based on their demand elasticity with the aim to design effective demand response programs.} 
	
	\highlight{When dealing with a customer pool consisting of different types of customers, the aggregated demand behavior is a combination of behaviors from distinctive energy user groups (e.g. C-HEMS, C-SM, and C-NONE considered in our paper) and will be very complicated.  For instance, there will be a lot of energy switch behaviors from C-HEMS and C-SM who are very sensitive to price changes (i.e. the demand would be an implicit discontinuous function of prices) whereas the demand of C-NONE is much less sensitive to price signals. Therefore, it is very difficult for existing demand modelling approaches which are mainly developed for a single type of users to handle the complicated demand behaviors of such a customer pool. To this end, in this paper we propose a hybrid demand modelling approach by considering the behavior differences among customers explicitly (i.e. customers of similar behavior patterns are categorized in the same group). Our proposed approach, which captures more detailed utility information of a mixed customer pool, can better reflect the demand behaviours in reality, and thus provide more accurate and well-behavioural demand models for the retailer to make right pricing decisions.}
	 
	 \highlight{In terms of bilevel optimization, there are many existing solution methods such as the commonly used single level reduction method which converts the bilevel problem into a single level problem by replacing the lower level optimization problem with its Karush--Kuhn--Tucker (KKT) conditions \cite{lu2016multilevel, sinha2017review}. However, for bilevel problems that have non-convex and discontinuous lower level problems such as our considered hybrid optimization+ machine learning problem with mixed (both integer and continuous) decision variables, the above conventional bilevel methods are infeasible \cite{ sinha2017review} due to the unavailability of derivative information and the non-convexity of lower level problems. In such cases, metaheuristic optimization methods such as genetic algorithms, which are easy to implement and do not require derivative information, are often employed and have been widely used in energy system modelling studies \cite{arabali2013genetic, lv2016bi, qu2016economic, trivedi2016genetic}. In addition, the intrinsic parallelism of genetic algorithms could be exploited in our future investigations e.g. by developing parallel genetic algorithms based solutions to take advantage of distributed and parallel computing facilities. To this end, in this paper we propose a GA-based two-level solution framework to solve the problem in a distributed and coordinated manner.} 

	 The main objective of this paper is to support the retailer to make best dynamic pricing decisions in such a mixed customer pool which takes all different types of customers (i.e., C-HEMS, C-SM, and C-NONE) into account at the same time. To the best of our knowledge, this is the first paper to tackle such a realistic and meaningful demand response problem by considering potential responses of different types of customers. The main contributions of this paper can be  summarized as follows:

	$\bullet$ We propose an integrated \textit{optimization+ learning} to optimal dynamic pricing for the retailer with a mixed customer pool by considering potential responses of different types of customers.
	
	$\bullet$  A genetic algorithm (GA) based two-level distributed pricing optimization framework is proposed for the retailer \highlight{to} determine optimal electricity prices.

	The rest of this paper is organized as follows. The system model framework is presented in Section \ref{Section-framework}.  An optimal home energy management system for C-HEMS is given in Section \ref{Section-HEMS} while two appliance-level learning models for C-SM are presented in Section \ref{Section-learning}. In Section \ref{Section-whole-learning}, an aggregated demand model for C-NONE is presented. A pricing optimization
	model for the retailer is provided in Section \ref{Section-pricing} and GA based two-level distributed pricing optimization algorithms are presented in Section \ref{Section-solutions}. Numerical results are presented in Section \ref{simulation_results}. This paper is concluded in Section \ref{conclusion}.

	\section{System Model} \label{Section-framework}
	
\begin{figure}[!t]				
					\centering
					\includegraphics[width=0.9\textwidth]{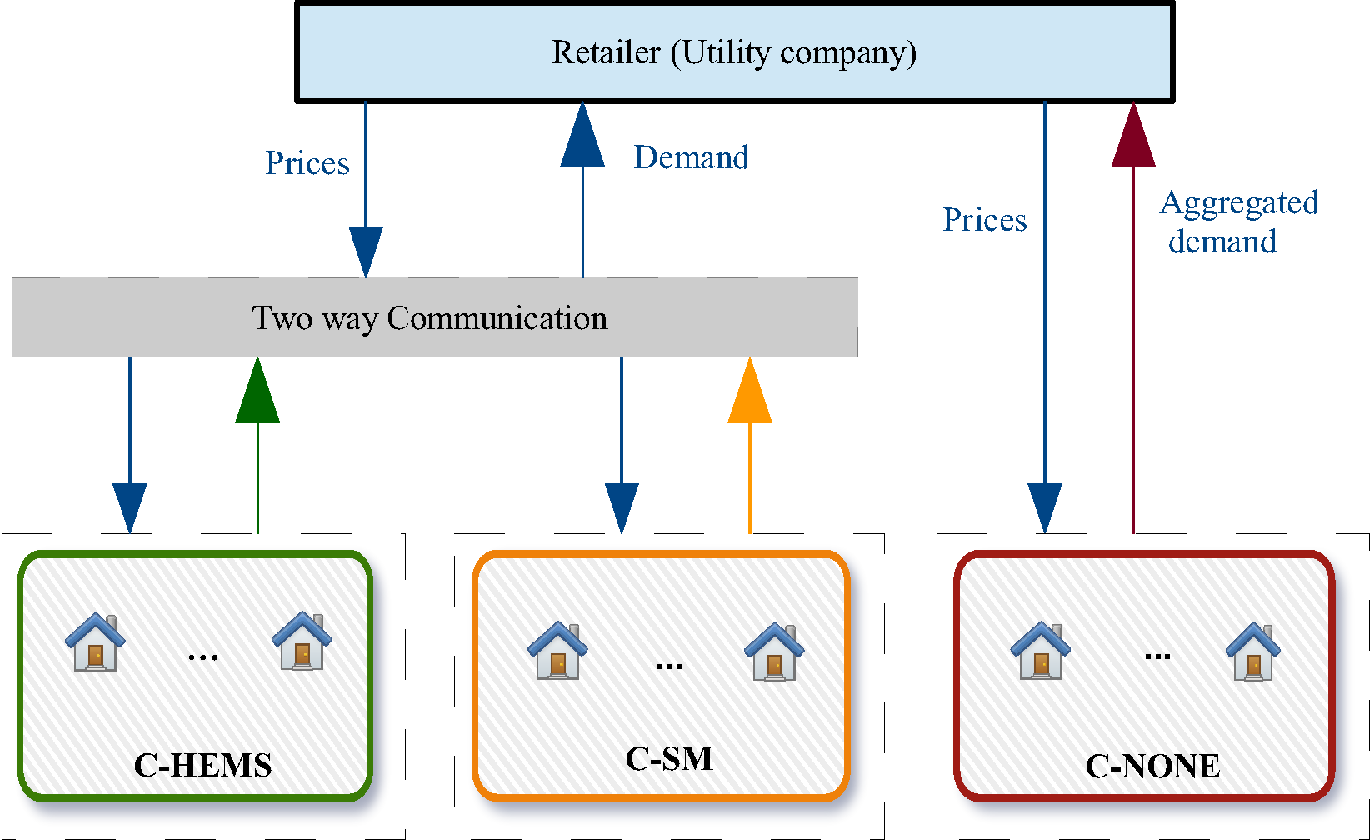}
					\caption{Two-level pricing optimization framework with the retailer and its customers.} 
					\label{figure:framework}
				\end{figure}

			In this paper, we consider a practical situation where a retailer serves three different groups of customers (i.e., C-HEMS, C-SM, and C-NONE). The number of above three groups of customers is denoted as $N_1$, $N_2$, and $N_3$ respectively with the total number of customers denoted as $N = N_1 + N_2 + N_3$.  The retailer procures electricity from the wholesale market, and then determines optimal retail dynamic prices based on the potential responses (when and how much is the energy consumption) of customers, which can be cast as a two-level decision making framework. The above interactions between the retailer and its customers is further depicted in Figure \ref{figure:framework}.

				As aforementioned, at the customer-side, for C-HEMS, their installed HEMS represents some optimization software such as to minimize customers' bills or maximize their comforts. With the help of two-way communication infrastructure, the retailer is able to know these customers' energy consumption responses to dynamic price signals by interacting with the installed HEMS.  As a result, for illustration purposes,  we formulate the energy management optimisation problem for C-HEMS by modifying \cite{meng2013stackelberg}. \highlight{In \cite{meng2013stackelberg}, we only consider shiftable and curtailable appliances where the scheduling problem of shiftable appliances is formulated as a linear program, and that of curtailable appliances is represented by a predefined linear demand function. However, in this paper we consider three types of appliances (i.e. interruptible, non-interruptible and curtailable appliances), with a more detailed problem formulation where the consumption scheduling problems of interruptible and non-interruptible appliances are formulated as integer programming problems, while that of curtailable appliances is formulated as a linear program. }  Nevertheless, other existing HEMS methods such as \cite{paterakis2015optimal}\cite{zhang2016model} in the literature should work equally well in this context. 
		
				For C-SM, they cannot always find the best energy usage scheduling without the help of HEMS. However, with the help of non-intrusive appliance load monitoring techniques (NILM) \cite{hart1992nonintrusive}, hourly or even minutely appliance-level energy consumption data of each appliance can be disaggregated from household-level smart meter data with high accuracy. As a result, appliance-level energy consumption patterns of each C-SM can be identified from history price and consumption data using machine learning algorithms. To this end, we modify \cite{meng2016profit} to identify appliance-level energy consumption patterns for each C-SM. \highlight{More specifically, in this paper we have removed the price elasticity of demand constraints considered in \cite{meng2016profit} for the learning model of curtailable appliances to simplify the model implementation, as such constraints seem to make no difference in this particular study.}
			
				For C-NONE, they usually manifest a relatively low demand elasticity due to lack of direct access to real-time price signals. On the other hand, the retailer is unable to know the accurate energy consumption information of each individual C-NONE. As a result, to identify the electricity consumption patterns for the pool of C-NONE, an aggregated demand model is needed. To this end, we \highlight{adopt} the approach proposed in \cite{ma2015demand} to identify the aggregated energy consumption patterns of the whole C-NONE. \highlight{Different from \cite{ma2015demand}, in this paper we have added a detailed analysis of the adopted aggregated demand model such as its capability of ensuring a basic market behavior and enabling the market operator or retailers to see the cross effect of usage switching.}

			 At the retailer-side, with the demand modelling for different types of consumers established, the pricing optimization problem for the retailer is formulated so as to maximize its profit. Such a two-level interaction model with hybrid  optimization (such as integer programming for C-HEMS) and machine learning problems (such as probabilistic Bayesian learning) at the customer-side and a quadratic programming problem at the retailer-side is non-convex and discontinuous. To this end, we propose a GA based solution framework to solve the retailer-side and customer-side problems in a distributed and coordinated manner. 
			 
			It is worth mentioning that we mainly focus on the pricing optimization problem from the perspective of retailer in this paper.  As a result, the benefits of customers are not discussed to depth in this specific study and readers are referred to our previously published works for more information \cite{meng2013stackelberg, meng2016profit} \cite{ma2015demand}. Furthermore, it should be noted that the obtained optimal dynamic prices will be applied to all types of customers.  A possible extension to the present work is to introduce a differential pricing framework for the retailer to offer individualized prices to each type of customers.  However, to determine the `right' pricing strategies for \highlight{different types of} customers \highlight{requires further} research and substantial numerical experiments, which is part of our future work.

\section{HEMS Optimization Model for C-HEMS} \label{Section-HEMS}

	In this section, we provide the  mathematical representation of the optimization model for customers with HEMS.  We define $\mathcal{N}_1 = \{ 1,2,...,N_1 \}$ as the set of such consumers.
	
	For each customer $n \in \mathcal{N}_1$, we denote the set of all the considered appliances as $A_n$, interruptible appliances (e.g., electric vehicles) as $I_n$, non-interruptible appliances (e.g., washing machine) as $NI_n$, and curtailable appliances (e.g., air conditioners) as $C_n$. As a result, we have $A_n = I_n \cup NI_n \cup C_n$. Since both interruptible and non-interruptible appliances can be regarded as shiftable appliances,  $S_n$ is used to represent such an union set, i.e.,  $S_n = I_n \cup NI_n$. 
	
	We define $\mathcal{H}= \{1,2,...,H\}$, where $H$ is the scheduling horizon for appliance operations. Further, let $p^h$ denote the electricity price announced by the retailer at $h \in \mathcal{H}$. For each appliance $ a \in A_n $, a scheduling vector of energy consumption over $H$ is denoted as $\mathrm{x}_{n,a}=[x_{n,a}^1,...,x_{n,a}^h,...,x_{n,a}^H]$
	where $ x^h_{n,a} \geq 0$ represents the $n$-th customer's energy consumption of appliance $a$ at time $h$. 
	
	\subsection{Interruptible Appliances}
	
	For each interruptible appliance $ a \in I_n $, the scheduling window for each appliance $a$ can be set by each customer according to his/her preference and is defined as $\mathcal{H}_{n,a} \triangleq \{\alpha_{n,a},  \alpha_{n,a}+1,...,\beta_{n,a}\}$. Note that the operations of these appliances can be discrete, i.e.,  it is possible to charge the EVs for one hour, then stop charging for one or several hours and then complete the charging after that. It is further assumed that the appliances consume a constant amount of energy (denoted as $x_{n,a}^{rated}$)  for each running time period.

	Finally, the payment minimization problem of each interruptible appliance is modelled as follows, which can be solved using existing integer linear programming solvers.
	\begin{align}
		&\min J_{I_n(a)} = \min_{x_{n,a}^h} \sum_{h=\alpha_{n,a}}^{\beta_{n,a}} p^h \times x_{n,a}^h \label{1}\\
		&s.t. \nonumber\\
		&\sum_{h=\alpha_{n,a}}^{\beta_{n,a}} x_{n,a}^h=E_{n,a}, \label{2}\\
		& x_{n,a}^h \in \{0, x_{n,a}^{rated}\}, \forall h \in
		\mathcal{H}_{n,a} \label{3}.
	\end{align}
	
	Constraint (\ref{2}) represents that, for each appliance $a$, the total energy consumption to accomplish the operations within the scheduling window is fixed, which \highlight{can be found from the technical specification of the appliance} and is denoted as $E_{n,a}$. Constraint (\ref{3}) represents that  appliance $a$ consumes a constant amount of energy for each running time period, i.e., $x_{n,a}^{rated}$. \highlight{Note that the actual running time period of an appliance $a$ can be derived from the vector of optimal energy consumptions, i.e. the appliance is `on' at hour $h$ if the corresponding optimal energy consumption is $x_{n,a}^{rated}$. Otherwise, the appliance is `off'.}

	\subsection{Non-interruptible Appliances}

	For each non-interruptible appliance  $a \in {NI_n}$, the scheduling window is defined as $\mathcal{H}_{n,a} \triangleq \{\alpha_{n,a}, \alpha_{n,a}+1,...,\beta_{n,a}\}$. Different from interruptible appliances, the operations of each non-interruptible appliance must be continuous, i.e., once the appliance starts, it must operate continuously till complete its task. Further, the appliances are assumed to consume a constant amount of energy (denoted as $x_{n,a}^{rated}$)  for each running time period.  The length of operations for each no-interruptible appliance is denoted as $L_{n,a}$.

	Finally, the bill minimization problem of each non-interruptible appliance is modelled as follows, which can be solved using existing integer linear programming solvers.
	\begin{align}
		& \min J_{NI_{n(a)}} = \min_{\delta_{n,a}^h} \sum_{h=\alpha_{n,a}}^{\beta_{n,a}} p^h \times  x_{n,a}^{rated} \times  \delta_{n,a}^h \label{5}\\
		& s.t. \nonumber \\
		&\sum_{h=\alpha_{n,a}}^{\beta_{n,a}} \delta_{n,a}^h = L_{n,a}, \label{6} \\
		& \delta_{n,a}^h  \in \{0,1\}, \forall h \in \mathcal{H}_{n,a}, \label{7} \\ 
		& \delta_{n,a}^{\alpha_{n,a} - 1 } = 0,  \label{8} \\ 
		&\nonumber \delta_{n,a}^h - \delta_{n,a}^{h-1} \leq  \delta_{n,a}^ \tau, \tau = h, ..., h + L_{n,a} - 1; \\ 
		& \hspace{6em}    \forall h \in \{\alpha_{n,a},...,\beta_{n,a} -L_{n,a} + 1\}. \label{9}
	\end{align} 
	
	Constraint \eqref{6} represents that the total time required to accomplish the operations of each non-interruptible appliance is predetermined and denoted as $L_{n,a}$. \eqref{7} indicates that the decision variable is a 0/1 variable representing the on/off operations of the appliance and \eqref{8} illustrates that the initial state of the appliance is `off'. \eqref{9} is adopted to guarantee the continuous operations of the appliance.
	
	\subsection{Curtailable Appliances} \label{subsec:curtailable}
	
		For each curtailable appliance $a \in {C_n}$, we define the scheduling window $ \mathcal{H}_{n,a} \triangleq \{\alpha_{n,a}, \alpha_{n,a} + 1,...,\beta_{n,a}\}$. The key characteristics which make curtailable appliances fundamentally different from interruptible and non-interruptible appliances are: 1) their energy demand cannot be postponed or shifted; 2) but their energy consumption level can be adjusted.

	In view of this, we define the energy consumption at time slot $h$ for each curtailable appliance $x_{n,a}^h$.  The minimum acceptable and maximum affordable consumption levels, which can be set in advance according to each individual customer's preferences, are defined as $\underline{u}_{n,a}^h$ and  $\overline{u}_{n,a}^h$ respectively. 

	Finally, the optimization problem of each curtailable appliance is proposed for each customer to minimize his/her payment bill subject to an acceptable total energy consumption, which can be solved via existing linear programming solvers.
	\begin{align}
		&{\operatorname{\min  }} J_{C_{n(a)}} = \underset{x_{n,a}^h} {\operatorname{\min}} \sum _{h=\alpha_{n,a}}^{\beta{n,a}} p^h \times x_{n,a}^h  \label{13} \\
		&s.t. \nonumber\\
		&\underline{u}_{n,a}^h   \leq x_{n,a}^h \leq \overline{u}_{n,a}^h, \label{14} \\
		&\sum_{h=\alpha_{n,a}}^{\beta{n,a}} x_{n,a}^h \geq U_{n,a}^{min}. \label{15}
	\end{align}
	
	Constraint (\ref{14}) enforces that the energy consumption at each time slot is within the minimum acceptable consumption level $\underline{u}_{n,a}^h$ and maximum affordable consumption level $\overline{u}_{n,a}^h$. Constraint (\ref{15}) indicates that for each curtailable appliance, there is a minimum acceptable total energy consumption during the whole operation periods that must be satisfied.

	\section{Appliance-level Learning Models for C-SM} \label{Section-learning}

	For each customer $n \in \mathcal{N}_2$, we define the set of shiftable appliances $S_n$ and curtailable appliances $C_n$. 
	
	For notation simplicity, we use $s$ to denote each shiftable appliance and $c$ for each curtailable appliance. Further, subscript $n$ is omitted in the rest of this section. 
	
	\subsection{Shiftable Appliances} \label{shiftable}
	
	Denote the scheduling window for each shiftable appliance $s$ as $\mathcal{H}_{s} \triangleq \{a_{s},...,b_{s}\}$,  where $a_{s}$ is the earliest possible time to switch on the appliance $s$ and $b_{s}$ is the latest possible time to switch off. Let $T_s = b_s - a_s + 1$ denote the length of the scheduling window for $s$. Assume the available historical smart data for appliance $s$ are electricity consumption scheduling vectors $\mathrm{x_{s}(d)} = [ x_{s}^{a_{s}}(d), x_{s}^{a_{s} + 1 }(d),...,x_{s}^{b_{s}}(d)]\; (d=1,2,...,D)$, where $x_{s}^{h}(d)\; (h = a_{s}, a_{s} +1,..., b_{s})$  represents the electricity consumption during time slot $h$ by appliance $s$ on day $d$. Suppose each shiftable appliance runs at a constant power rate, and the total running time and electricity taken for appliance $s$ to accomplish the operations are denoted as $L_s$ and  $E_{s}$ respectively.

Based on the above historical data of a given customer showing when appliance $s$ has been used and the corresponding dynamic prices, the basic idea behind this appliance-level learning model is to calculate the probabilities that appliance $s$ was used at the cheapest, second cheapest,..., or most expensive price. The above insights can be represented as \eqref{recursive_initial}.
\begin{equation} \label{recursive_initial}
P_{i}^{s}(d)=\frac{f_{i}(d)}{d} \quad d=1,2,...D
\end{equation} 
where $f_{i}(d)$ represents the number of days when appliance $s$ is used i-th cheapest within the past $d$ days. Note that the superscript $s$ at the right hand side of the equation has been omitted for notation simplicity. Let the current day be $d$ and then $f_{i}(d)$ and $P_{i}^{s}(d) $ can be derived based on  historical data up to day $d$ and \eqref{recursive_initial} respectively. 

Further, let $\delta_{i} (d+1)$ (taking value as 1 or 0) represent the probability that appliance $s$ is used i-th cheapest on day $d+1$ and then it becomes a new piece of information to be used to obtain $P_{i}^{s} (d+1)$. As a result, \eqref{recursive_initial} can be rewritten in a recursive way as follows:  \begin{equation} \label{recursive_1} 
\begin{array}{l} 
P_{i}^{s} (d+1)=\frac{f_{i} (d+1)}{d+1} \\ {=P_{i}^{s} (d)+\frac{1}{d+1} [\delta _{i} (d+1)-P_{i}^{s} (d)]}.
\end{array} 
\end{equation}
	
The above recursive formula shows that when a new piece of information $\delta_{i} (d+1)$ is received, the updated probability, $P_{i}^{s} (d+1)$, is equal to the existing probability $P_{i}^{s} (d)$ plus an adjusting term. The adjusting term includes the adjusting coefficient $1/(d+1)$ and the prediction error term $[\delta_{i} (d+1)-P_{i}^{s} (d)]$. Recall that $\delta _{i} (d)$ only takes its value as 1 or 0 in \eqref{recursive_1}, which also means the cost (sum of hourly electricity prices) of each possible operation schedule for appliance $s$ is different with each other. However, under some circumstances, many hourly prices are the same within a day and possibly for many days, which could result in two or more operation schedules having the same costs such that there are more than one $i$-th cheapest operation schedules. To overcome such uncertainties in the price signals, a systematic framework for obtaining $\delta_{i} (d+1)$ under different cases is to be given below. 
	
Firstly, suppose that there are in total $k$ possible operation schedules for appliance $s$ and the cost of each schedule (i.e. sum of the hourly prices) on day $d+1$ is $c_{j} (d+1)$ where $j=1,...,k$ is an unique index for each schedule. Secondly, $c_{j}(d+1)$ is sorted in an ascending order. For the cases where there are two or more schedules having the same cost, these costs are treated as the order they appear in  $c_{j}(d+1) \; (j=1,...,k)$. As a result, the cost of $m$-th cheapest schedule can be denoted as $r_{m} (d+1)\; (m=1,...,k)$.

	Finally, when electricity prices and the usage data of each appliance are received at the end of day $d+1$, $\delta _{i} (d+1)$ is calculated based on the following three cases. 
	
	$\bullet$  Case 1. $s$ is not operated as the i-th cheapest schedule on day $d+1$, then 
	$$\delta_{i} (d+1)=0;$$ 
	
	$\bullet$  Case 2. $s$ is operated using the i-th cheapest schedule on day $d+1$ with the cost $r_{i} (d+1)$ which satisfies $r_{i} (d+1) \ne r_{m}(d+1)\; (\forall m,\; m\ne i)$, then 
	$$\delta_{i} (d+1)=1;$$
	
	$\bullet$ Case 3. $s$ is operated using the i-th cheapest schedule on day $d+1$ with the cost $r_{i} (d+1)\; $ but there are $k_{0}$ additional  operation schedules where  $r_{m_{l} } (d+1)=r_{i} (d+1)\; (l=1,...,k_{0})$, then 
	$$\delta_{i} (d+1)=\frac{P_{i}^{s} (d)}{P_{i}^{s} (d)+\sum _{l=1}^{k_{0} }P_{m_{l} }^{s} (d) }.$$

	Based on the above probabilistic usage model $P_{i}^{s}$, the expected bill of each shiftable appliance $s$ on a given day can be calculated from the perspective of retailer as follows. 
	\begin{equation}\label{bill}
	B_s =  \sum\limits_{i} \begin{pmatrix} PC_i^s \times E_s/L_s \times P_i^s \end{pmatrix}
	\end{equation} where $PC_i^s$ denote the cost of i-th cheapest schedule $PT_{i}^{s}$ on that given day. 
	
	Furthermore, the expected hourly energy consumption of each shiftable appliance $s$ can be calculated as follows.
	\begin{equation}\label{hourly consumption}
	\overline{y_{s,h}} = \sum\limits_{i} \begin{pmatrix} E_s/L_s  \times P_i^s \times I_{i,s}^h \end{pmatrix}
	\end{equation} where $I_{i,s}^h$ is defined as follows:
	\begin{equation} \label{indicator_function}
	\nonumber	I_{i,s}^h =
	\begin{cases} 
	1 & \text{if } h \in PT_{i}^{s}  \\ 
	0 & \text{if } h \notin PT_{i}^{s}
	\end{cases}.
	\end{equation}

	\subsection{Curtailable Appliances} \label{Curtailable Appliances}

	The scheduling window of appliance $c \in C_n$ is defined as $\mathcal{H}_{c} \triangleq \{a_{c},...,b_{c}\}$. Let $ \big\{y_{c,h}(d), \bar{\textbf{p}}(d) \big \}=  \big \{y_{c,h}(d), [p^{a_{c}}(d), ..., p^{b_c}(d)] \big \}, d= 1,...,D $ be the available historical input-output data of an unknown demand function, where the input data $\bar{\textbf{p}}=[ p^{a_{c}}(d),..., \\ p^{b_c}(d)]$ represent the price signals during the scheduling window $\mathcal{H}_c$ on day $d$ and the output data $y_{c,h}(d)$ represent the energy consumption of appliance $c$ at time slot $h$ on day $d$. 

	We use a linear demand function to model how a customer responds to dynamic price signals when using curtailable appliances, which is formulated as follows.
	\begin{equation}\label{curtailable demand}
	\hat{y}_{c,h} = \alpha_{c,h,0} + \beta_{c,h,a_{c}}p^{a_{c}} + ...+ \beta_{c,h,b_c}p^{b_c}
	\end{equation}
	where $\hat{y}_{c,h}$ represents the expected demand of appliance $c$ at time $h$, $p^{h}$ is the electricity price at time slot  $h$,  $\alpha_{c,h,0}, \beta_{c,h,a_{c}},..., \beta_{c,h,b_c}$ are the parameters to be identified.
	
	In this paper, the least square is adopted to estimate model parameters $\boldsymbol{\beta} = [\alpha_{c,h,0}, \beta_{c,h,a_{c}},\\ ..., \beta_{c,h,b_c} ]$, where the best estimates $\hat {\boldsymbol{\beta}}$ are obtained by solving \eqref{qp}.
		\begin{equation} \label{qp}
		\begin{array}{cc}
		\hat {\boldsymbol{\beta}} = \underset{\boldsymbol{\beta}} {\operatorname{arg min}} \; \sum_{d=1}^{D} \big( y_{c,h}(d) - \alpha_{c,h,0} -  \beta_{c,h,a_c} p_{a_c}(d) - ... -  \beta_{c,h,b_c} p_{b_c}(d) \big)^2.
		\end{array}
		\end{equation}
		
	Finally, the expected hourly energy consumption of appliance $c$ for a given day, denoted as $\overline{y_{c,h}}$, can  be predicted based on the above established demand model. Further, the expected daily bill of $c$ on that day can be represented as $B_{c}= \sum\limits_{h \in \mathcal{H}_{c}} p^h \times \overline{y_{c,h}}.$

	\section{Aggregated Demand Modelling for C-NONE} \label{Section-whole-learning}
	
	In this section, an aggregated demand model is proposed to identify the demand patterns of C-NONE.
	
	Suppose the electricity prices and aggregated consumption data for all C-NONE in last $D$ days are available. We define the electricity price vector on day $d \in \mathcal{D} \triangleq \{1,..., D\}$ as ${\textbf{p}(d)}=\left[ p^1(d),...,p^h(d),...,p^H(d) \right]$ where $p^h(d)$ represents the price at hour $h \in \mathcal{H}$ on day $d$. Furthermore, we define the aggregated electricity consumption vector on day $d \in \mathcal{D}$ as $\textbf{y}(d) = \left[ y_1(d),...,y_h(d),...,y_H(d)\right]$, where $y_h(d)$ represents the aggregated consumption by all C-NONE at hour $h$.
	
It is believed that the aggregated electricity demand of C-NONE at hour $h$ not only depends on the price at $h$ but also on prices at other hours due to the cross effect of usage switching  \cite{kirschen2000factoring} \cite{ma2015demand}. Therefore the aggregated demand model at hour $h$ can be expressed as follows:
	\begin{equation} \label{demand_real_function}
	y_h = R_h(p^1,p^2,...,p^H).
	\end{equation}
		
As the mathematical form of $R_h(\cdot)$ is usually unknown, we need to find an estimated demand function $\hat{R}_h(p^1, p^2,...,p^H)$ that is as close to $R_h(p^1,p^2,...,p^H)$ as possible. For this purpose, we use a linear demand function to represent $\hat{R}_h(p^1,p^2,...,p^H)$ as follows.
\begin{equation} \label{lineardemand}
\begin{array}{cccccc}
\hat{R}_h(p^1, p^2,...,p^H) = \alpha_h +  \beta_{h,1}p^1 + \dots + \beta_{h,h}p^h  + \dots + \beta_{h,H}p^H.
\end{array}
\end{equation}
where $\beta_{h,h}$ is called the self or direct price elasticity of demand, which measures the responsiveness of electricity demand at hour $h$ to changes in the electricity price at $h$. 	\highlight{When the price at hour $h$ increases but prices at other times remain unchanged, the demand at hour $h$ typically decreases, and thus the self-elasticity is usually negative \cite{kirschen2000factoring} (see \eqref{elasity1})}. $\beta_{h,l} \;(h \neq l)$ is the cross-price elasticity, which measures the responsiveness of the demand for the electricity at hour $h$ to changes in price at some other hour $l \neq h$. \highlight{When the price of electricity at hour  $l$ increases but prices at other times remain unchanged, some demand at hour $l$ will typically be shifted to hour $h$, and therefore the demand at hour $h$ increases. Thus, cross elasticities are usually positive \cite{kirschen2000factoring} (see \eqref{elasity2})}. Furthermore, we consider an important necessary and sufficient condition (see \eqref{elasity3}) for the electricity to be a demand consistent retail product to ensure that the proposed demand model follows a normal market behavior. That is, when the overall market price is decreased, the overall market demand should increase or remain unchanged. 
\begin{equation} \label{elasity1}
\beta_{h,h}  < 0.
\end{equation}
\begin{equation} \label{elasity2}
\beta_{h,l} > 0 \mbox{ if }  h \neq l.
\end{equation}
\begin{equation} \label{elasity3}
\beta_{h,h} +  \sum_{l \in \mathcal{H}, l \neq h} \beta_{l,h} \leq 0
\end{equation}

It should be emphasized that the important reasons behind using linear demand models are: 1) As the prices of electricity
are normally changing slowly with time, at a given time, we only need to model the demand around a small price interval or locally. Since any non-linear behavior can be approximated well via a  linear model locally, it is the main reason that linear demand model is widely used in this research area \cite{ma2015demand} \cite{aalami2010modeling} and  selected in this work; 2) with conditions \eqref{elasity1} - \eqref{elasity3}, it ensures the basic market behavior that demands go down when prices go up and vice versa during the pricing optimization. However, nonlinear demand models often fail to maintain this basic market rule and could result in situations such as higher market prices leading to higher usages. As a result, using such nonlinear demand models in the pricing optimization could result in incorrect pricing for the retailer;  3) it enables the market operator or retailers to see the cross effect of usage switching such as customers usually only shift their usages to nearby hours but rarely to far away hours.

Finally, the demand model parameters $\beta_{h,l}, l = 1,...,H$ can be identified by solving the following optimization problem. 
\begin{equation} \label{qp_whole_demand}
\begin{array}{lll}
 \qquad     \min \; \sum_{d=1}^{D} \sum_{h=1}^{H} \lambda^{(D-d)}\big(  \alpha_{h} +  \beta_{h,1} p^{1}(d)  +  \dots +   \beta_{h,H} p^{H}(d) - y_h(d) \big)^2 \\
\mbox{\qquad  subject to \eqref{elasity1}, \eqref{elasity2}, and \eqref{elasity3}.}
\end{array}
\end{equation} where $0 \le \lambda \leq 1 $ is the forgetting factor which exponentially discounts the influence of old data and therefore the model will catch up the behavior change of customers with time. \eqref{qp_whole_demand} is a quadratic programming problem and can be solved using existing solvers.

	\section{Pricing Optimization for the Retailer} \label{Section-pricing}
	
	From this section, we restart using the subscript $n$ in all of the following mathematical representations, which has been omitted in Section \ref{Section-learning}.

	We define a cost function $C_h(L_h)$ to represent the retailer's cost of providing $L_h$ electricity to all customers at each hour $h \in \mathcal{H}$. We make the same assumption as \cite{mohsenian2010autonomous} that the cost function $ C_h(L_h)$ is convex increasing in $ L_h $, which is designed as follows.
	\begin{equation} \label{costfunction}
	C_h(L_h) = a_h L_h^2 + b_h L_h + c_h 
	\end{equation}
	where $a_h > 0$ and $b_h \geq 0, c_h \geq 0$ for each hour $h \in \mathcal{H}$.

	We denote the minimum price (e.g., the wholesale price) that the retailer can offer as $p_h^{min}$ and the maximum price (e.g., the retail price cap due to retail market competition and regulation) as $p_h^{max}$. As a result, we have:
	\begin{equation}\label{20}
	p_h^{min} \leq p^h \leq p_h^{max}.
	\end{equation}
	
	A maximum energy supply at each time slot, denoted as  $E_h^{\max }$, is imposed on the retailer to respect the power network capacity. Thus, we have
	\begin{equation}\label{21}
	\begin{array}{cl}
	E_h = \sum\limits_{n \in \mathcal{N}_1} \sum\limits_{a \in A_n} {x_{n,a}^h}    + \sum\limits_{n \in \mathcal{N}_2} \big ({\sum\limits_{s \in {S_n}} {\overline{y_{n,s,h}}}} + {\sum\limits_{c \in {C_n}}{\overline{y_{n,c,h}}}} \big)  +  \hat{R}_h(p^1, p^2,...,p^H) \le E_h^{\max}, \forall h \in \mathcal{H} 
	\end{array}
	\end{equation}

	Due to the retail market regulation, we add the revenue constraint to ensure a sufficient number of low price periods and thus to improve the acceptability of retailer's pricing strategies. That is, there exists a total revenue cap, denoted as $RE^{max}$, for the retailer. Thus, we have the following constraint \footnote{It should be highlighted that such a revenue cap is necessary for the pricing optimization. Since the electricity is the basic necessity in daily life and fundamentally less elastic, without such a revenue cap, a retailer could lift its profit significantly by increasing its prices aggressively. However, such a pricing strategy will anger customers and could lead to political consequences \cite{freezeenergyprices}. For such a reason, the revenue cap which basically is the total customers' bill cap is necessary to ensure the sensible pricing strategy for a retailer.}:
	\begin{equation}\label{22}
	\begin{array}{clll}
	RE= \sum \limits_{h \in \mathcal{H}} p^h \times \sum\limits_{n \in \mathcal{N}_1} {\sum\limits_{a \in A_n} x_{n,a}^h}  +  \sum\limits_{n \in \mathcal{N}_2} ( \sum\limits_{s \in {S_n}} B_{n,s} + \\  \sum\limits_{c \in {C_n}} B_{n,c} )  + \sum \limits_{h \in \mathcal{H}} p^h \times \hat{R}_h(p^1, p^2,...,p^H)  \leq RE^{max}
	\end{array}
	\end{equation}

	Finally, the profit maximization problem for the retailer to optimize the electricity prices for the next day is modelled as follows:
	\begin{equation} \label{profit maximization1}
	\begin{array}{ccc}
	\underset{p^h} {\operatorname{\max}} \left\lbrace RE -  \sum \limits_{ h\in \mathcal{H}} C_h(E_h)\right\rbrace
	\\
	\mbox{\qquad  subject to constraints (\ref{20}), (\ref{21}), and (\ref{22}).}
	\end{array}
	\end{equation}

	\section{GA based Solution Algorithms to the above Two-level Model}
	 \label{Section-solutions}
	 
	 \subsection{Solution Framework}
	
	As the proposed two-level pricing optimization problem consisting of the profit maximization problem for the retailer and integrated \textit{optimization+ learning} based demand modelling problems for customers is non-convex, non-differentiable and discontinuous, it is intractable for conventional non-linear optimization methods. As a result, we adopt a genetic algorithm (GA) based solution methods to solve the problems for retailer and its customers in a distributed and coordinated manner.


	In our proposed genetic algorithms, binary encoding and deterministic tournament selection without replacement is adopted. For the crossover and mutation operations, we employ uniform crossover and bit flip mutation respectively. The constraints are handled by the approach proposed in \cite{deb2000efficient}. Readers are referred to \cite{meng2013stackelberg} for more details on our adopted GA.
	
	Finally, the GA based distributed pricing optimization framework are given in Algorithm \ref{Algorithm:GA based decision-making scheme algorithm}. Moreover, the optimal home energy management algorithm for C-HEMS is given in Algorithm \ref{Energy scheduling algorithm}. The appliance-level learning algorithms for C-SM is presented in Algorithm \ref{Learning algorithm}. For C-NONE, the retailer could directly use the established demand models presented in Section \ref{Section-whole-learning} to forecast customers' consumptions. It is worth mentioning that at the end of each day, the established demand models of C-SM and C-NONE will be updated based on the newly available prices and usages data of that day.
	At the end, the optimal dynamic prices are found for the retailer.
	
		\begin{algorithm}[!ht]
			
			\renewcommand{\algorithmicrequire}{\textbf{}}
			
			\caption{GA based pricing optimization algorithm to \eqref{profit maximization1} executed by the retailer}
			\label{Algorithm:GA based decision-making scheme algorithm}
			
			\begin{algorithmic}[1]
				
				\STATE Population initialization by generating a population of $PN$ chromosomes randomly; each chromosome represents a strategy (i.e. prices over $H$) of the retailer.
				
				\FOR{ i=1 to $PN$ } 
				\STATE The retailer announces strategy $i$ to customers.
				
				\STATE Receive the responsive demand from $n \in \mathcal{N}_1 \cup \mathcal{N}_2 $ (i.e., Algorithm \ref{Energy scheduling algorithm} and \ref{Learning algorithm}). In addition, the responsive demands from C-NONE are estimated based on the aggregated demand model proposed in Section \ref{Section-whole-learning}.
				
				\STATE Fitness evaluation and constraint handling \cite{deb2000efficient} to satisfy constraints (\ref{20} - \ref{22}). 
				\ENDFOR
				\STATE A new generation are created by using deterministic tournament selection without replacement, uniform crossover and bit flip mutation.
				
				\STATE Steps 2-7 are repeated until the stopping condition is reached and the retailer announces final prices to all the customers.
			\end{algorithmic}
		\end{algorithm}
		
		\begin{algorithm}[!ht]
			\renewcommand{\algorithmicrequire}{\textbf{}}
			\caption{HEMS executed by each smart meter for C-HEMS}
			\label{Energy scheduling algorithm}
			\begin{algorithmic}[1]
				\STATE Receive the price signals from the retailer via smart meter.
				\STATE The smart meter schedules energy consumption based on prices by solving optimization problems in Section \ref{Section-HEMS}.
				\STATE The smart meter only sends back the aggregated hourly demand of the household to retailer via the two-way communication infrastructure. 
			\end{algorithmic}
		\end{algorithm}

		\begin{algorithm}[!ht]
			\renewcommand{\algorithmicrequire}{\textbf{}}
			\caption{Appliance-level learning executed by each smart meter for C-SM}
			\label{Learning algorithm}
			\begin{algorithmic}[1]
				
				\STATE Receive the price signals from the retailer via smart meter.
				
				\STATE  The smart meter calculates the expected hourly energy consumption and daily bill payment of each appliance based on the learning models proposed in Section \ref{Section-learning}.
				\STATE The smart meter sends back aggregated hourly  demand and daily bill information of the household to the retailer via the two-way communication infrastructure. 
			\end{algorithmic}
		\end{algorithm} 
	
	\subsection{Computational Aspects of the Model}
	
 \highlight{The considered bilevel model has a hierarchical structure where the retailer acting as the upper level agent and customers acting as decentralized lower level agents (for C-NONE, the problem is solved directly at the retailer side). The subsequent proposed solution framework solves the problem in a distributed and coordinated manner, where the retailer determines the prices first and then customers (C-HEMS and C-SM) simultaneously determines their consumptions based on the price signal. Since each customer (\highlightR{C-HEMS} and C-SM) is equipped with a smart meter which is assumed to have the computational capacity to solve its own consumption scheduling problem independently and simultaneously when receiving the price signal, it is believed that increasing the customer number does not increase the total computation time significantly. Similar conclusions have been reported in \cite{qian2013demand} where a simulated annealing algorithm is adopted to solve their upper level pricing optimization problem with independent customers. In addition, a similar solution procedure as ours where the upper level problem is firstly evaluated via metaheuristic optimization algorithms and the lower level problem is tackled via standard solvers (e.g., integer programming and linear programming solvers) has been recently reported in \cite{han2016solution} and been proved to be effective in solving large scale bi-level problems.}

\highlight{Although promising, it should be noted that bilevel optimization problems are generally difficult to solve (e.g., for the simplest case where both upper and lower level problems are linear programs, it is proved to be NP-hard \cite{ben1990computational}) and might face scalability issues in solving very large scale problems.  One possible solution to overcome the scalability issues is to find  good initial solutions for the problem by utilizing and learning from historical data using machine learning algorithms, which can therefore greatly reduce the number of iterations needed. For instance, in our considered bilevel problem, due to the daily pricing practice, there are many historical data of past daily prices and customer responses (consumptions). By going through these data at the retailer side, some approximated optimal prices can be found, which can be chosen as starting points for GA. As a result, the optimal prices for the next day are likely to be found within a smaller number of iterations.}

	\section{Simulation Results} \label{simulation_results}

	In this section, we conduct simulations to evaluate the proposed pricing optimization model with different types of customers. \highlight{Ideally, we would use observed data from relevant trials in the simulations, however, not all of them are available. As a result, for those data which are publicly available (e.g., the electricity price and demand data used for the aggregated demand modelling \cite{newengland2013}), relevant links/references have been cited in the text. For those data which are not publicly available, we will  simulate the required data (e.g., the dynamic electricity price and consumption response data at the appliance level used for customer behaviour learning model for C-SM) or provide the required data directly in the paper (e.g., the parameter settings of appliances for the C-HEMS models).}
				
	\subsection{Simulation Set-up}
	We simulate a neighbourhood consisting of one retailer and  three different types of customers (i.e., C-HEMS, C-SM, C-NONE) where the total number of customers is set to 100.  It is assumed that each customer has 5 appliances: EV, dishwasher, washing machine, clothes dryer and air conditioner. Further, a fixed amount (0.05 kWh) of background consumption at each hour is considered for each household. The scheduling window is set from 8AM to 8AM (the next day). It is worth mentioning that customers are assumed to be homogeneous due to hardware constraints with the aim to simply the model implementation process. For details on how to implement a similar kind of model with heterogeneous customers, readers are referred to \cite{qian2013demand}.
%
	
		\begin{table}[!t]
			\caption{Parameters for each interruptible appliance}
			\label{Table:1} 
			\centering 
			\begin{tabular}{llll}
				\hline
				Appliance Name & $E_{a}$  & $H_{a}$ &  $x_{n,a}^{rated}$ \\ 
				\hline\noalign{\smallskip} 
				
				Dishwasher   & \texttt{1.8kWh}   & \texttt{8PM-7AM} & \texttt{ 1kWh} \\
				PHEV   & \texttt{10kWh}   & \texttt{7PM-7AM} & \texttt{2.5kWh}
				\\
				\hline
			\end{tabular}
		\end{table}
		
		\begin{table}[!t]
			\caption{Parameters for each non-interruptible appliance}
			\label{Table:2} 
			\centering 
			\begin{tabular}{lllll}
				\hline
				Appliance Name & $E_{a}$  & $H_{a}$ & $x_{n,a}^{rated}$ & $L_a$ \\
				\hline
				Washing machine  & \texttt{2kWh}   & \texttt{8AM-9PM} & \texttt{1kWh} & \texttt{2hrs} \\
				Clothes Dryer  & \texttt{3kWh}   & \texttt{8PM-6AM} & \texttt{1.5kWh} & \texttt{2hrs} \\
				\hline
			\end{tabular}
			
		\end{table}

		\begin{table}[!t]
			\caption{Parameters for each curtailable appliance}
			\label{Table:3} 
			\centering 
			\begin{tabular}{lllll}
				\hline
				Appliance Name & $ U_{a}^{min}$  & $H_{a}$ &  $ \underline{u}_{a}^h $ & $\overline{u}_{a}^h$ \\
				\hline\noalign{\smallskip}
				Air-conditioner  & \texttt{18kWh}  & \texttt{12PM-12AM} & \texttt{1kWh} & \texttt{2kWh} \\
				\hline
			\end{tabular}
			
		\end{table}

	The parameter settings for HEMS optimization models of  C-HEMS are given in Tables \ref{Table:1} \ref{Table:2} \ref{Table:3}.	\highlight{Despite an extensive search, however, we have not found any publicly available real-world data for real-time prices and consequence demand response at the individual appliance level, data which are required for our simulation for C-SM.} As a result, the historical usage data used in appliance-level customer behavior learning for shiftable appliances in C-SM are generated by tuning HEMS optimization models with waiting time costs \cite{meng2016profit} whereas the historical usage data for identifying the demand model of curtailable appliances are simulated based on Section \ref{subsec:curtailable}. For C-NONE, the aggregated demand model is learned from history electricity price data and down-scaled energy consumption data (i.e., daily consumption of each household is scaled down to the same amount as C-HEMS and C-SM) between 1 January 2012 and 21 December 2012 of ISO New England \cite{newengland2013}. For the retailer, the maximum price and minimum price at each hour are set to 6.00 cents and 14.00 cents respectively. \highlight{Further, the parameters of GA are set as Table \ref{Table:Parameter settings of GA}. More specifically, the chromosome length $L_g$  is determined based on the value range (i.e. [6.00 cents, 14.00 cents] in our study) and precision requirement (i.e. two digits after the decimal point) of each decision variable $p^h, h=1,2,...24$. Since there are in total $(14.00-6.00)/0.01 = 800$ possible values for each variable $p^h$, at least 10 binary bits ($2^{9} <800$ and $2^{10} >800$) are needed to satisfy the precision requirement of each variable. Therefore, the chromosome, which represents a vector of 24 decision variables, needs to be (at least) 240 binary bits long \cite{meng2013stackelberg}. To set the mutation rate $P_g$, we start with the standard mutation rate setting (i.e. $1/L_g \approx  0.0042$ in our study), and finally find that a mutation rate of 0.005 works well for our problem. For the population size, which usually increases with the problem size (dimensions), we start with the setting used in \cite{deb2000efficient} (i.e. the population size is set to 10 multiplied by the problem dimension -- 240 in our study), and finally find that a population size of 300 works well for our problem. Considering the problem complexity, the termination generation is set to 300 to ensure sufficient evolutions of the GA.}
%
			
		\begin{table}[!t]
			\caption{Parameter settings of GA}
			\label{Table:Parameter settings of GA}  
			\centering 
			\begin{tabular}{lll}
				\hline
				Parameter Name & Symbol  & Values\\
				Chromosome Length & $L_g$  & \highlight{10 $\times$ 24}\\
				Population Size   & $PN$  & 300 \\
				Mutation Probability   & $P_g$ & 0.005 \\
				Termination Generation & $ T_g$  &  300 \\
				\hline
			\end{tabular}
		\end{table}

	\subsection{Results and Analysis}

			\highlight{The simulations are implemented in the Matlab R2016b software environment on a 64 bit Linux PC with Intel Core i5-6500T CPU @ 2.50GHz and 16 GB RAM.}
			
\begin{figure}[!t]
		\centering
		\includegraphics[width=0.8\textwidth]{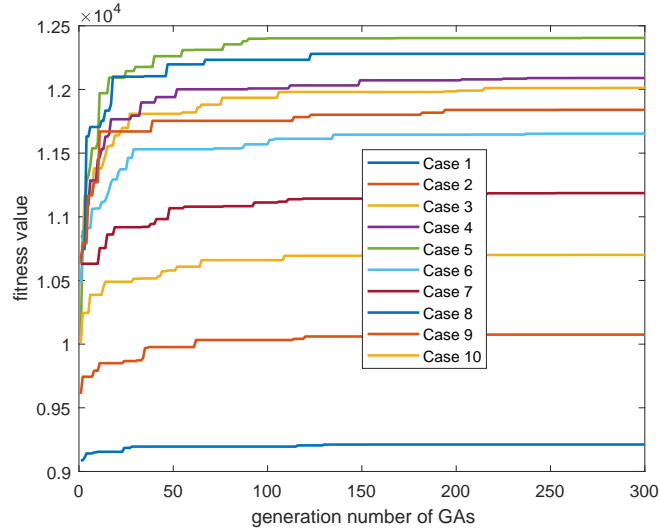}
		\caption{The evolution process of GA with different penetration rates of smart meter and HEMS.} 
		\label{figure:SM-HEMS-None}
	\end{figure}

	We implement the proposed pricing optimization under each case study listed in Table \ref{Table:casestudy} where the evolution processes of GA are shown in Figure \ref{figure:SM-HEMS-None}. The revenue, cost and profit of the retailer \highlight{as well as the CPU computation time} under different case studies are given in Table \ref{Table:result1}. \highlight{Note that for case studies where C-HEMS and C-SM coexist (C-NONE are directly handled at the retailer side), the energy consumption problems of  C-HEMS and C-SM under each retailer's pricing strategy are solved sequentially in this paper, which leads to a CPU time of around 2000 seconds. However, since in real applications customers' energy consumption problems are tackled simultaneously by each smart meter in parallel, the total computation time is only constrained by the customers energy consumption problem with the highest computation time and will be much shorter (i.e. around 1000 seconds for C-HEMS in this study). It is also worth mentioning that our proposed optimal dynamic pricing for the retailer is updated every 24 hours, and therefore the solution method is sufficiently fast for this purpose.}

	\begin{table}[!t]
		\caption{Combinations of different types of customers for pricing optimization}
		\label{Table:casestudy} 
		\centering
		\begin{tabular}{llll}
			\hline
			Case study number & C-HEMS & C-SM & C-NONE \\
			\hline\noalign{\smallskip}  
			1   & 0 & 0	& 100 	 \\ 
			2    & 0 &  30	&  70 \\
			3    & 0 &  100	&  0 \\
			4    & 30 &  70	&  0 \\
			5    & 100 &  0	&  0 \\
			6    & 50 &  30	&  20 \\
			\highlight{	7} & 	0 &  	70	&  	30 \\
			\highlight{		8  } & 70 &  30	&  0 \\
			\highlight{		9 }  & 30 &  60	&  10 \\
			\highlight{		10}  & 20 &  30	&  50 \\
			\hline
		\end{tabular}  
	\end{table}
	
	\begin{table}[!t]
		\caption{Comparison of revenue, cost and profit of the retailer as well as the CPU computation time under different case studies}
		\label{Table:result1} 
		\centering
		\begin{tabular}{llllllll}
			\hline
			Case study number & Revenue (\$) & Cost (\$) & Profit (\$) & \highlight{CPU time  (s) }\\
			\hline\noalign{\smallskip}      
			1  & 350 &   257.88
			&   92.12 & 9	 \\
			2  & 350 &  249.25
			&  100.75	& 838 \\
			3   & 350 & 229.88
			&  120.12	& 853 \\
			4   & 350 & 228.89
			& 121.11	 & 2015\\
			5   & 350 & 225.95
			& 124.05	 & 1013 \\
			6   & 350 & 233.47
			&   116.53 & 1999 \\
			\highlight{7}   & 350 & 238.14
			&  111.86	& 838  \\
			\highlight{ 8}   & 350 & 227.20
			& 122.80	& 2013 \\
			\highlight{ 9}   & 350 & 231.60
			& 118.40	& 2010 \\
			\highlight{10}   & 350 & 242.99
			&   107.01 & 1997 \\
			\hline
		\end{tabular}  
	\end{table}

	In addition, it can be found from Table \ref{Table:result1} that, if all customers are C-SM, the profit of retailer will be \$120.12 compared with that of \$92.12 under 100\% C-NONE.  In addition, the profit of retailer can be further increased to \$124.05 if all customers become C-HEMS. The above results show that, without increasing its revenue, the retailer can actually gain more profit with the installation of smart meters and HEMS in households, which indicates that the retailers are likely to have strong motivations to participate in and promote the proposed pricing based demand response programs.

\begin{figure}[!ht]	
				\centering
				\includegraphics[width=0.75\textwidth]{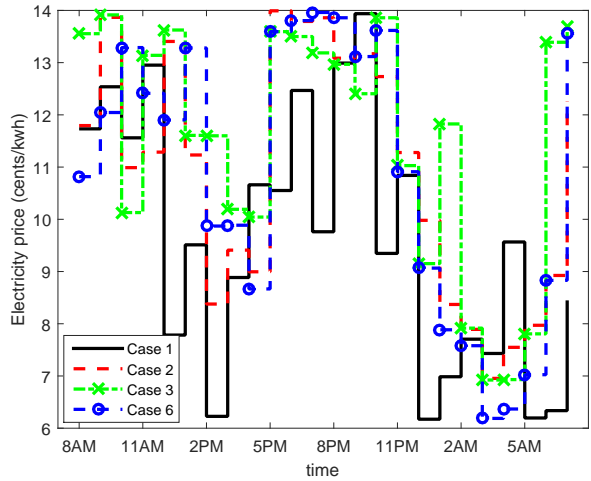}
				\caption{Optimal prices for different case studies.} 
				\label{figure:prices}
			\end{figure}

\begin{figure}[!ht]	
				\centering
				\includegraphics[width=0.75\textwidth]{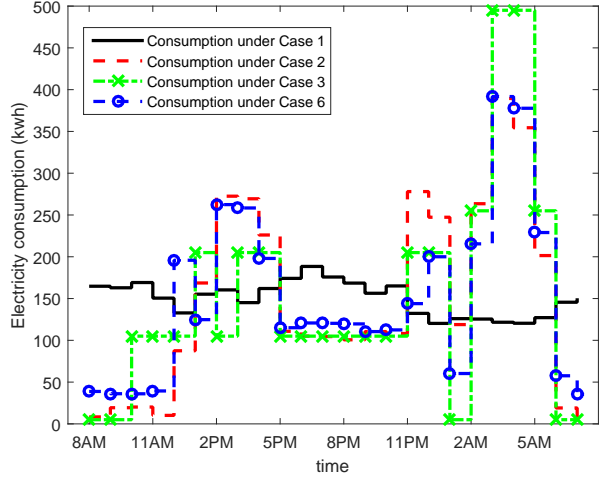}
				\caption{Electricity demand of customers under  different case studies.} 
				\label{figure:demand}
			\end{figure}

		Furthermore, the optimized electricity prices under several selected case studies are shown in Figure \ref{figure:prices} where the corresponding energy consumption of customers are plotted in Figure \ref{figure:demand}. It can be found from Figures \ref{figure:prices} \ref{figure:demand} that, compared with that of C-NONE, the demands of C-HEMS and SM are more responsive to prices, and are effectively shifted from peak-demand periods (higher prices) to off-peak demand periods (lower prices). In the case where 3 different types of customers co-exist, the resulted prices and demands exhibit a mixed and synthesized behavior. For instance, the demand of a mixed pool of customers under Case 6 is still responsive to price signals but less sensitive than that of C-HEMS (Case 3). In other words, in a realistic scenario like Case 6, our proposed pricing optimization model can generate profitable electricity prices for the retailer and  lead to a reasonable consumption behavior for customers.

	\subsection{Emerging Scenarios with PV Generation and Energy Storage}

	In the following,  we consider the availability of PV generation for all types of customers (e.g., C-HEMS, C-SM, C-NONE) and also battery energy storage for C-HEMS. The pricing optimization for the retailer under such  emerging scenarios are investigated in this subsection. 
	
	Suppose that a forecast of PV generation is known as a prior at the beginning of each day. It is further assumed that PV generation will be firstly used by customers whereas any surplus will be sold back to the grid at same \highlight{retail} price of that time. The PV generation data used in this simulation is adopted from \cite{zhang2016model}. Furthermore,  the revenue cap imposed on the retailer is reduced accordingly from \$350 to \$270 due to the availability of PV generation and therefore the reduction of energy demand of customers. In addition, we assume that C-HEMS use energy storage for price arbitrage to minimize their energy cost. The energy storage model is adopted from \cite{zhang2016model} and we consider \textit{perfect charging and discharging}, i.e., the charging/discharging efficiency factor is equal to $1$. The battery capacity is considered as 10 kWh and the maximum amount of energy to be charged/discharged in each time period is constrained to 2 kWh. Furthermore, the initial and final state of charge (SoC) are set to 80\% (i.e., 8 kWh in this simulation).  Note that a full battery energy storage model considering self-discharging and degradation \cite{vetter2005ageing} is out of scope of this paper. 

	\begin{table}[!t]
		\caption{Comparison of revenue, cost and profit of the retailer under different case studies with PV generation}
		\label{Table:result4} 
		\centering
		\begin{tabular}{llllllll}
			\hline
			Case study number & Revenue (\$) & Cost (\$) & Profit (\$) \\
			\hline\noalign{\smallskip}      
			1  & 270  &    198.64
			&    71.36	 \\
		    3  & 270 &  170.66
			&  99.34	 \\
			5   & 270 & 167.29
			&  102.71	 \\
			6   & 270 & 174.63
			& 95.37	 \\
			\hline
		\end{tabular}  
	\end{table}

			\begin{table}[!t]
				\caption{Comparison of revenue, cost and profit of the retailer under C-HEMS with PV and energy storage}
				\label{Table:result5} 
				\centering
				\begin{tabular}{lllll}
					\hline
					Scenario & Revenue (\$) & Cost (\$) & Profit (\$) \\
					\hline\noalign{\smallskip} 
					
					C-HEMS with PV and storage (no selling back) & 270 & 145.45& 124.55	 \\
					
					C-HEMS with PV and storage (selling back)  & 270  & 124.98 & 145.02	 \\
					
					\hline
				\end{tabular}  
			\end{table}
				
The impact of PV penetration on the retailer is investigated under case studies shown in Table \ref{Table:casestudy}, where the simulation results can be found in Table \ref{Table:result4}. Compared with those in Table \ref{Table:result1}, we can see that the retailer achieves similar profit levels (i.e., ratio of profit to revenue). The above finding indicates that the profit of retailer is not significantly influenced by the penetration of PV generation.

\begin{figure}[h]
	\centering
	\includegraphics[width=0.7\textwidth]{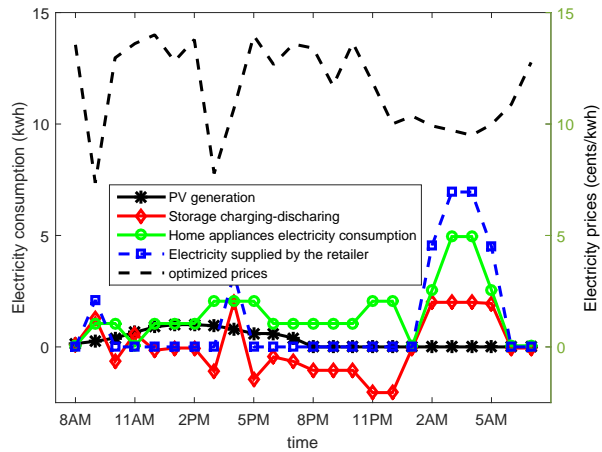}
	\caption{Storage operation and appliance consumption profile of one household (no selling back).} 
	\label{figure:Storage-nosellback}	
\end{figure}

\begin{figure}[h]
	
	\centering
	\includegraphics[width=0.7\textwidth]{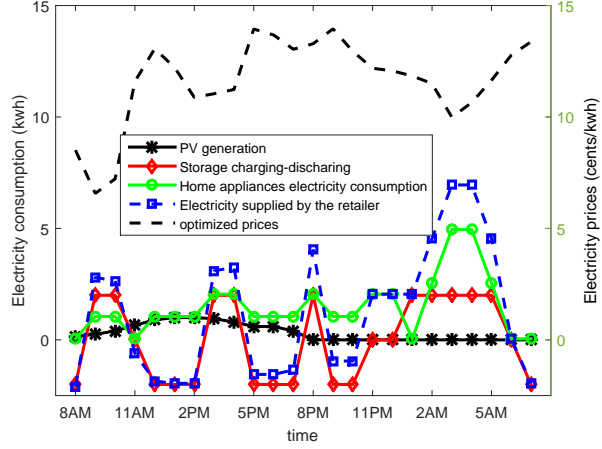}
	\caption{Storage operation and appliance consumption profile of one household (selling back).} 
	\label{figure:Storage-sellback}	
\end{figure}

Additionally, we study the impact of battery energy storage on the retailer where customers are C-HEMS.  We consider two scenarios for C-HEMS (i.e.,  with/without the capability of selling electricity back to the grid) where the profits of retailer are given in Table \ref{Table:result5}. The above results reveal two-fold findings: 1) the retailer can improve its profit with the energy storage penetration in households; 2) the retailer could gain even more profit if customers (C-HEMS) have the capability of selling electricity back to the grid. \highlight{To be more specific, with battery energy storage, customers will have the demand flexibility where some electricity can be bought at cheap price time periods beforehand and stored in the energy storage units for use in high price periods. As dynamic prices reflect wholesale prices, the purchase cost of the retailer for such electricity in the wholesale market will also be low. Without increasing its revenue, the retailer can increase its profit. In addition, if customers can sell electricity back to the grid, the energy storage charging/discharging operations are no longer constrained by the amount of electricity customers actually use (i.e. the energy storage units charge/discharge at higher power rates), which gives customers more demand flexibility. Thus,  more electricity can be bought at cheap price time periods beforehand, which on the other hand leads to a lower wholesale electricity purchase cost for the retailer. As a result, without increasing its revenue, the retailer can gain even more profit.} Furthermore, the optimized prices and corresponding storage operations and appliance consumption profiles in the above two scenarios are illustrated in Figures  \ref{figure:Storage-nosellback} and  \ref{figure:Storage-sellback}.  It can be easily found out that the customers with capacity of selling back electricity often  charge/discharge the battery at maximum rate in order to take full advantage of price arbitrage to minimize their energy bills.

\subsection{\highlightR{Solution Algorithms Comparison}}

\highlightR{In general, algorithms to solve bilevel problems can be approximated categorized into three groups: iterative algorithms based on the definitions of bilevel/Stackelberg equilibriums (e.g.,\cite{mediwaththe2017competitive}), KKT based classical mathematical optimization algorithms (i.e. KKT based single level reduction) and GA type of metaheuristic algorithms \cite{sinha2017review}. In this subsection, we conduct further simulations to compare our GA based solution algorithm with the other two types of algorithms.}

\highlightR{Firstly, we compare GA with a two-step iterative algorithm  \cite{mediwaththe2017competitive} on selected representative case studies listed in Table \ref{Table:casestudy}. Same as our GA based solution method, the iterative algorithm is also working in a distributed manner that the retailer and customers solve their optimization problems sequentially. The description of the algorithm in the context of our problem setting is given as follows: (1) find an initial feasible price vector for the retailer; (2)
given a price vector announced by the retailer, each customer solves their energy consumption problem and obtains optimal energy consumption in response to the price vector; (3) the retailer receives energy consumptions  of all customers and treats them as known parameters in the profit maximization model Eq. \eqref{profit maximization1}. Therefore,  Eq. \eqref{profit maximization1} becomes a linear programming problem, from solving which the retailer obtains a new price vector; (4) If the previous two price vectors satisfy the termination criterion (i.e. prices are close enough), the best solution is recorded and the algorithm is terminated; otherwise, go to step (2). }

\highlightR{The solutions obtained by GA and the iterative algorithm are reported in Table \ref{Table:comparison1}. Note that the reported results of the iterative algorithm are based on the best initial conditions among several attempts for each case study. From the results, we can find that GA outperforms the iterative algorithm in all cases. It is also observed that the iterative algorithm performs worst in the case where customers are all C-HEMS (mathematical optimization models at the lower level) whereas it performs best in the case where customers are all C-NONE (analytical electricity demand functions at the lower level). The above can explained by the fact that the iterative algorithm is a local search based heuristic and is more likely to be trapped in the local minima, especially for more complex problems. Instead, GA is a population based meta-heuristic algorithm with global search characteristics and has more chances to find the global optimal/near-optimal solutions.}

\begin{table}[!t]
	\caption{Comparison of GA with the iterative algorithm}
	\label{Table:comparison1} 
	\centering
	\begin{tabular}{lllll}
		\hline
		\multirow{2}{*}{Case study number} & \multicolumn{2}{l}{  \;\; \; \; \; \;  GA} & \multicolumn{2}{l}{Iterative algorithm}    \\ \cline{2-5} 
		   & Revenue &   Profit  &    Revenue &  Profit  \\ \hline
		1  &  350    &   \textbf{92.12}   &    350 &  89.98   \\ 
      	3  &  350    &   \textbf{120.12}  &    350  &  80.05   \\ 
    	5 &  350     &   \textbf{124.05}  &     350  &  73.48    \\
		6  & 350     &   \textbf{116.53}  &    350  &   80.50   \\      
		\hline
	\end{tabular}
\end{table}

\highlightR{Secondly, we are also interested in knowing how good solutions that our proposed GA could achieve compared with theoretical optimal or near-optimal solutions, which are usually obtained by classical mathematical optimization methods such as KKT based single level reduction method. It should be noted that the KKT based method solves the bilevel model in a centralized manner at the retailer's side, by which it assumes that an optimistic bilevel model is adopted, i.e. customers are always expected to make decisions that lead to the best possible profit of the retailer \cite{sinha2017review}. Clearly, the above assumption is strong and therefore the best solutions obtained by the KKT based centralized method (given that the real optimums are attained) are deemed to be better than solutions obtained by  distributed solution methods considering game behaviours between the retailer and customers such as our GA based distributed method. Since the KKT based method requires a well-defined convex and continuous lower level optimization problem, it cannot be  directly used to solve our bilevel model. To make the comparisons achievable, we modify our bilevel model such that only C-HEMS are considered at the lower level problem. In addition, the integer linear programming problems of interruptible and non-interruptible appliances of C-HEMS are modified to linear programs. Finally, the modified bilevel model has a lower level with only linear programming problems and the KKT based method is implemented in Matlab with the assistance of YALMIP toolbox \cite{lofberg2004yalmip}.}

\highlightR{In this particular simulation, we consider six different model instances (customer profiles) of the relaxed C-HEMS model, which are named from  C-HEMS-1 to C-HEMS-6 respectively. Except for the first customer profile (C-HEMS-1) which is directly adapted from Tables \ref{Table:1} \ref{Table:2} \ref{Table:3}, the other five customer profiles are generated by adding random numbers to the above settings. Finally, we compare GA with KKT based method on six test problems as shown in Table \ref{comparison_2} . For all the test problems, GA uses the same settings as Table \ref{Table:Parameter settings of GA}. For KKT based method, we set the relative optimality gap of solvers to 0.01\% and also a large amount of computation time to achieve as best solutions as possible to provide a reliable benchmark. }

\begin{table}[!t]
	\caption{Comparison of GA with KKT based single level reduction method}
		\label{comparison_2}
	\begin{tabular}{llcccc}
		\hline
		\multirow{2}{*}{Test problem} & \multirow{2}{*}{Customer profiles \tablefootnote{The numbers 1 to 6 represents C-HEMS-1, C-HEMS-2, ... C-HEMS-6 respectively.}} & \multirow{2}{*}{\makecell{Number of lower \\ level decision variables}} & GA & \multicolumn{2}{l}{\; \;\;\;\;\;KKT}   \\ \cline{4-6} 
		&                                         &         & Profit & Profit&  Optimality gap  \\ \hline
		1     & 1     & 63     & \textbf{124.05}  & \textbf{124.05} & 0.0000\%       \\
		2     & 1- 2  & 131   & 96.76  & \textbf{96.80} & 0.0000\%        \\
		3    & 1- 3     & 221   & 127.47 & \textbf{127.53} & 0.0036\%       \\
		4    & 1- 4    & 296    & \textbf{135.98} & 135.96 & 0.0281\%       \\
		5   & 1- 5    & 351    & \textbf{141.62} & 141.58 & 0.0344\%      \\
		6   & 1- 6 & 460     & \textbf{134.19}   & 133.74 & 0.4142\%      \\
		\hline
	\end{tabular}
\end{table}

\highlightR{The simulation results are reported in Table \ref{comparison_2}, from which we can find that for relatively small-scale test problems such as problems 1-3, KKT based centralized method could achieve the theoretic optimums or very close optimums with a small optimality gap. On the other side, GA based distributed solution method can attain the theoretic optimum for test problem 1 and could also achieve solutions close to theoretic optimums for test problems 2-3. For larger-scale problems such as test problems 4-6, KKT based centralized method could not achieve an optimality gap of 0.01\% within reasonably large computation times (the computation time limit is set to 12000 seconds) and only near optimums are attained. It is also observed that our GA based solution method could achieve actually better near optimums than KKT based solution method on these large scale test problems. As a result, from the above comparisons of GA with the iterative algorithm and KKT based single level reduction method on different test cases, it is reasonable to conclude that our proposed GA are good enough and efficient in solving the proposed bilevel model. }

\highlightR{It is also worth pointing out that the proposed GA based distributed solution method is also a feasible and effective solution for the realistic cases of energy pricing problem where there may be up to a few millions of customers. The reason is that, by simply distributing the computing of all customers' best responses to the retailer's prices to a small number (saying a few hundreds) of cloud computing devices in which each device computes the best responses of a group of customers (saying a couple of thousands), the corresponding cloud/parallel computing enables the time needed for computing all customers' best responses being largely similar to the time needed for computing of each group of customers. In other words, the proposed GA based solution method enables the parallel computing and leads to the feasibility and effectiveness for the realistic cases of energy pricing with a large number of customers. This is in contrast with the complexity of KKT based optimization algorithms in the same cases, for which it requires the centralised computing with multi-millions of constraints in order to find the feasible and optimal prices. As for the definition based iterative algorithms, the local optimization feature of such algorithms makes it much weaker option compared with our proposed approach.}

	\section{Conclusion} \label{conclusion}
	
	In this paper, we study the dynamic pricing optimization problem in a realistic scenario consisting of one retailer and three different types of customers (C-NONE, C-SM, and C-HEMS). The interactions between retailer and customers are treated as a two-level decision-making framework. Firstly, we propose an integrated \textit{optimization+ machine learning} based demand modelling framework for customers. Secondly, we propose a profit-maximization based dynamic pricing model for the retailer subject to realistic market constraints. Finally, GA based distributed pricing optimization algorithms are proposed to tackle the above two-level decision making problems.  Simulation results indicate that our proposed pricing optimization model and solution algorithms are feasible and effective. \highlightR{However, to understand the computational aspect of bilevel problems and their solution algorithms thoroughly, much work remains to be done. In our future work, we plan a separate, dedicated contribution on that subject considering both our and more generalized bilevel problems, by taking advantage of e.g., machine learning algorithms and distributed and parallel computational facilities.}

\section*{Acknowledgements}
	
		This work was partly supported by the National Nature Science Foundation of China (Grant No. 71301133),  Humanity and Social Science Youth Foundation of Ministry of Education, China (Grant No. 13YJC630033), and the Engineering and Physical Sciences Research Council, UK (Grant No. EP/I031650/1).

\section*{References}

	\bibliographystyle{elsarticle-num-names-alpha} 
	 \bibliography{INS}

\begin{thebibliography}{49}
\providecommand{\natexlab}[1]{#1}
\providecommand{\url}[1]{\texttt{#1}}
\providecommand{\urlprefix}{URL }
\expandafter\ifx\csname urlstyle\endcsname\relax
  \providecommand{\doi}[1]{doi:\discretionary{}{}{}#1}\else
  \providecommand{\doi}[1]{doi:\discretionary{}{}{}\begingroup
  \urlstyle{rm}\url{#1}\endgroup}\fi
\providecommand{\bibinfo}[2]{#2}

\bibitem[{fre(2016)}]{freezeenergyprices}
\bibinfo{howpublished}{\url{http://www.bbc.co.uk/news/uk-politics-24213366}},
  \bibinfo{year}{(accessed August 30, 2016)}.

\bibitem[{UKS(2017)}]{UKSM2017}
\bibinfo{howpublished}{\url{https://www.ofgem.gov.uk/electricity/retail-market/metering/transition-smart-meters}},
  \bibinfo{year}{(accessed July 16, 2017)}.

\bibitem[{new(2013)}]{newengland2013}
\bibinfo{howpublished}{\url{http://iso-ne.com/markets/hstdata/znl\_info/hourly/index.html}},
  \bibinfo{year}{(accessed March 01, 2013)}.

\bibitem[{Aalami et~al.(2010)Aalami, Moghaddam, and
  Yousefi}]{aalami2010modeling}
\bibinfo{author}{H.~Aalami}, \bibinfo{author}{M.~P. Moghaddam},
  \bibinfo{author}{G.~Yousefi}, \bibinfo{title}{Modeling and prioritizing
  demand response programs in power markets}, \bibinfo{journal}{Electric Power
  Systems Research} \bibinfo{volume}{80}~(\bibinfo{number}{4})
  (\bibinfo{year}{2010}) \bibinfo{pages}{426--435}.

\bibitem[{Adika and Wang(2014)}]{adika2013autonomous}
\bibinfo{author}{C.~O. Adika}, \bibinfo{author}{L.~Wang},
  \bibinfo{title}{Autonomous appliance scheduling for household energy
  management}, \bibinfo{journal}{IEEE Transactions on Smart Grid}
  \bibinfo{volume}{5}~(\bibinfo{number}{2}) (\bibinfo{year}{2014})
  \bibinfo{pages}{673--682}.

\bibitem[{Arabali et~al.(2013)Arabali, Ghofrani, Etezadi-Amoli, Fadali, and
  Baghzouz}]{arabali2013genetic}
\bibinfo{author}{A.~Arabali}, \bibinfo{author}{M.~Ghofrani},
  \bibinfo{author}{M.~Etezadi-Amoli}, \bibinfo{author}{M.~S. Fadali},
  \bibinfo{author}{Y.~Baghzouz}, \bibinfo{title}{Genetic-algorithm-based
  optimization approach for energy management}, \bibinfo{journal}{IEEE
  Transactions on Power Delivery} \bibinfo{volume}{28}~(\bibinfo{number}{1})
  (\bibinfo{year}{2013}) \bibinfo{pages}{162--170}.

\bibitem[{Asadinejad et~al.(2016)Asadinejad, Varzaneh, Tomsovic, Chen, and
  Sawhney}]{asadinejad2016residential}
\bibinfo{author}{A.~Asadinejad}, \bibinfo{author}{M.~G. Varzaneh},
  \bibinfo{author}{K.~Tomsovic}, \bibinfo{author}{C.-f. Chen},
  \bibinfo{author}{R.~Sawhney}, \bibinfo{title}{Residential customers
  elasticity estimation and clustering based on their contribution at incentive
  based demand response}, in: \bibinfo{booktitle}{Power and Energy Society
  General Meeting (PESGM), 2016}, \bibinfo{organization}{IEEE},
  \bibinfo{pages}{1--5}, \bibinfo{year}{2016}.

\bibitem[{Ben-Ayed and Blair(1990)}]{ben1990computational}
\bibinfo{author}{O.~Ben-Ayed}, \bibinfo{author}{C.~E. Blair},
  \bibinfo{title}{Computational difficulties of bilevel linear programming},
  \bibinfo{journal}{Operations Research}
  \bibinfo{volume}{38}~(\bibinfo{number}{3}) (\bibinfo{year}{1990})
  \bibinfo{pages}{556--560}.

\bibitem[{Chai et~al.(2014)Chai, Chen, Yang, and Zhang}]{chai2014demand}
\bibinfo{author}{B.~Chai}, \bibinfo{author}{J.~Chen},
  \bibinfo{author}{Z.~Yang}, \bibinfo{author}{Y.~Zhang}, \bibinfo{title}{Demand
  response management with multiple utility companies: A two-level game
  approach}, \bibinfo{journal}{IEEE Transactions on Smart Grid}
  \bibinfo{volume}{5}~(\bibinfo{number}{2}) (\bibinfo{year}{2014})
  \bibinfo{pages}{722--731}.

\bibitem[{Datchanamoorthy et~al.(2011)Datchanamoorthy, Kumar, Ozturk, and
  Lee}]{datchanamoorthy2011optimal}
\bibinfo{author}{S.~Datchanamoorthy}, \bibinfo{author}{S.~Kumar},
  \bibinfo{author}{Y.~Ozturk}, \bibinfo{author}{G.~Lee},
  \bibinfo{title}{Optimal time-of-use pricing for residential load control},
  in: \bibinfo{booktitle}{Smart Grid Communications (SmartGridComm), 2011 IEEE
  International Conference on}, \bibinfo{organization}{IEEE},
  \bibinfo{pages}{375--380}, \bibinfo{year}{2011}.

\bibitem[{Deb(2000)}]{deb2000efficient}
\bibinfo{author}{K.~Deb}, \bibinfo{title}{An efficient constraint handling
  method for genetic algorithms}, \bibinfo{journal}{Computer Methods in Applied
  Mechanics and Engineering} \bibinfo{volume}{186}~(\bibinfo{number}{2})
  (\bibinfo{year}{2000}) \bibinfo{pages}{311--338}.

\bibitem[{Esther and Kumar(2016)}]{esther2016survey}
\bibinfo{author}{B.~P. Esther}, \bibinfo{author}{K.~S. Kumar},
  \bibinfo{title}{A survey on residential Demand Side Management architecture,
  approaches, optimization models and methods}, \bibinfo{journal}{Renewable and
  Sustainable Energy Reviews} \bibinfo{volume}{59} (\bibinfo{year}{2016})
  \bibinfo{pages}{342--351}.

\bibitem[{Flath et~al.(2012)Flath, Nicolay, Conte, van Dinther, and
  Filipova-Neumann}]{flath2012cluster}
\bibinfo{author}{C.~Flath}, \bibinfo{author}{D.~Nicolay},
  \bibinfo{author}{T.~Conte}, \bibinfo{author}{C.~van Dinther},
  \bibinfo{author}{L.~Filipova-Neumann}, \bibinfo{title}{Cluster Analysis of
  Smart Metering Data-An Implementation in Practice},
  \bibinfo{journal}{{Business \& Information} Systems Engineering}
  \bibinfo{volume}{4}~(\bibinfo{number}{1}) (\bibinfo{year}{2012})
  \bibinfo{pages}{31--39}.

\bibitem[{Fu et~al.(2017)Fu, Zeng, Luo, Wang, Xu, and Fan}]{fu2017designing}
\bibinfo{author}{X.~Fu}, \bibinfo{author}{X.-J. Zeng}, \bibinfo{author}{X.~R.
  Luo}, \bibinfo{author}{D.~Wang}, \bibinfo{author}{D.~Xu},
  \bibinfo{author}{Q.-L. Fan}, \bibinfo{title}{Designing an intelligent
  decision support system for effective negotiation pricing: A systematic and
  learning approach}, \bibinfo{journal}{Decision Support Systems}
  \bibinfo{volume}{96} (\bibinfo{year}{2017}) \bibinfo{pages}{49--66}.

\bibitem[{G\'{o}mez et~al.(2012)G\'{o}mez, Chertkov, Backhaus, and
  Kappen}]{gomez2012learning}
\bibinfo{author}{V.~G\'{o}mez}, \bibinfo{author}{M.~Chertkov},
  \bibinfo{author}{S.~Backhaus}, \bibinfo{author}{H.~J. Kappen},
  \bibinfo{title}{Learning price-elasticity of smart consumers in power
  distribution systems}, in: \bibinfo{booktitle}{Smart Grid Communications
  (SmartGridComm), 2012 IEEE Third International Conference on},
  \bibinfo{organization}{IEEE}, \bibinfo{pages}{647--652},
  \bibinfo{year}{2012}.

\bibitem[{Han et~al.(2016)Han, Zhang, Hu, and Lu}]{han2016solution}
\bibinfo{author}{J.~Han}, \bibinfo{author}{G.~Zhang}, \bibinfo{author}{Y.~Hu},
  \bibinfo{author}{J.~Lu}, \bibinfo{title}{A solution to bi/tri-level
  programming problems using particle swarm optimization},
  \bibinfo{journal}{Information Sciences} \bibinfo{volume}{370}
  (\bibinfo{year}{2016}) \bibinfo{pages}{519--537}.

\bibitem[{Hart(1992)}]{hart1992nonintrusive}
\bibinfo{author}{G.~W. Hart}, \bibinfo{title}{Nonintrusive appliance load
  monitoring}, \bibinfo{journal}{Proceedings of the IEEE}
  \bibinfo{volume}{80}~(\bibinfo{number}{12}) (\bibinfo{year}{1992})
  \bibinfo{pages}{1870--1891}.

\bibitem[{Herter(2007)}]{herter2007residential}
\bibinfo{author}{K.~Herter}, \bibinfo{title}{Residential implementation of
  critical-peak pricing of electricity}, \bibinfo{journal}{Energy Policy}
  \bibinfo{volume}{35}~(\bibinfo{number}{4}) (\bibinfo{year}{2007})
  \bibinfo{pages}{2121--2130}.

\bibitem[{Hosking et~al.(2013)Hosking, Natarajan, Ghosh, Subramanian, and
  Zhang}]{hosking2013short}
\bibinfo{author}{J.~Hosking}, \bibinfo{author}{R.~Natarajan},
  \bibinfo{author}{S.~Ghosh}, \bibinfo{author}{S.~Subramanian},
  \bibinfo{author}{X.~Zhang}, \bibinfo{title}{Short-term forecasting of the
  daily load curve for residential electricity usage in the Smart Grid},
  \bibinfo{journal}{Applied Stochastic Models in Business and Industry}
  \bibinfo{volume}{29}~(\bibinfo{number}{6}) (\bibinfo{year}{2013})
  \bibinfo{pages}{604--620}.

\bibitem[{Khan et~al.(2016)Khan, Mahmood, Safdar, Khan, and
  Khan}]{khan2016load}
\bibinfo{author}{A.~R. Khan}, \bibinfo{author}{A.~Mahmood},
  \bibinfo{author}{A.~Safdar}, \bibinfo{author}{Z.~A. Khan},
  \bibinfo{author}{N.~A. Khan}, \bibinfo{title}{Load forecasting, dynamic
  pricing and DSM in smart grid: A review}, \bibinfo{journal}{Renewable and
  Sustainable Energy Reviews} \bibinfo{volume}{54} (\bibinfo{year}{2016})
  \bibinfo{pages}{1311--1322}.

\bibitem[{Khotanzad et~al.(2002)Khotanzad, Zhou, and
  Elragal}]{khotanzad2002neuro}
\bibinfo{author}{A.~Khotanzad}, \bibinfo{author}{E.~Zhou},
  \bibinfo{author}{H.~Elragal}, \bibinfo{title}{A neuro-fuzzy approach to
  short-term load forecasting in a price-sensitive environment},
  \bibinfo{journal}{IEEE Transactions on Power Systems}
  \bibinfo{volume}{17}~(\bibinfo{number}{4}) (\bibinfo{year}{2002})
  \bibinfo{pages}{1273--1282}.

\bibitem[{Kirschen et~al.(2000)Kirschen, Strbac, Cumperayot, and
  de~Paiva~Mendes}]{kirschen2000factoring}
\bibinfo{author}{D.~S. Kirschen}, \bibinfo{author}{G.~Strbac},
  \bibinfo{author}{P.~Cumperayot}, \bibinfo{author}{D.~de~Paiva~Mendes},
  \bibinfo{title}{Factoring the elasticity of demand in electricity prices},
  \bibinfo{journal}{IEEE Transactions on Power Systems}
  \bibinfo{volume}{15}~(\bibinfo{number}{2}) (\bibinfo{year}{2000})
  \bibinfo{pages}{612--617}.

\bibitem[{Lofberg(2004)}]{lofberg2004yalmip}
\bibinfo{author}{J.~Lofberg}, \bibinfo{title}{YALMIP: A toolbox for modeling
  and optimization in MATLAB}, in: \bibinfo{booktitle}{Computer Aided Control
  Systems Design, 2004 IEEE International Symposium on},
  \bibinfo{organization}{IEEE}, \bibinfo{pages}{284--289},
  \bibinfo{year}{2004}.

\bibitem[{Lu et~al.(2016)Lu, Han, Hu, and Zhang}]{lu2016multilevel}
\bibinfo{author}{J.~Lu}, \bibinfo{author}{J.~Han}, \bibinfo{author}{Y.~Hu},
  \bibinfo{author}{G.~Zhang}, \bibinfo{title}{Multilevel decision-making: a
  survey}, \bibinfo{journal}{Information Sciences} \bibinfo{volume}{346}
  (\bibinfo{year}{2016}) \bibinfo{pages}{463--487}.

\bibitem[{Lv et~al.(2016)Lv, Ai, and Zhao}]{lv2016bi}
\bibinfo{author}{T.~Lv}, \bibinfo{author}{Q.~Ai}, \bibinfo{author}{Y.~Zhao},
  \bibinfo{title}{A bi-level multi-objective optimal operation of
  grid-connected microgrids}, \bibinfo{journal}{Electric Power Systems
  Research} \bibinfo{volume}{131} (\bibinfo{year}{2016})
  \bibinfo{pages}{60--70}.

\bibitem[{Ma and Zeng(2015)}]{ma2015demand}
\bibinfo{author}{Q.~Ma}, \bibinfo{author}{X.-J. Zeng}, \bibinfo{title}{Demand
  modelling in electricity market with day-ahead dynamic pricing}, in:
  \bibinfo{booktitle}{2015 IEEE International Conference on Smart Grid
  Communications (SmartGridComm)}, \bibinfo{organization}{IEEE},
  \bibinfo{pages}{97--102}, \bibinfo{year}{2015}.

\bibitem[{Mediwaththe et~al.(2017)Mediwaththe, Stephens, Smith, and
  Mahanti}]{mediwaththe2017competitive}
\bibinfo{author}{C.~Mediwaththe}, \bibinfo{author}{E.~Stephens},
  \bibinfo{author}{D.~Smith}, \bibinfo{author}{A.~Mahanti},
  \bibinfo{title}{Competitive Energy Trading Framework for Demand-side
  Management in Neighborhood Area Networks}, \bibinfo{journal}{IEEE
  Transactions on Smart Grid}  (\bibinfo{year}{2017}) \bibinfo{pages}{(in
  press)}.

\bibitem[{Meng and Zeng(2013)}]{meng2013stackelberg}
\bibinfo{author}{F.-L. Meng}, \bibinfo{author}{X.-J. Zeng}, \bibinfo{title}{A
  Stackelberg game-theoretic approach to optimal real-time pricing for the
  smart grid}, \bibinfo{journal}{Soft Computing}
  \bibinfo{volume}{17}~(\bibinfo{number}{12}) (\bibinfo{year}{2013})
  \bibinfo{pages}{2365--2380}.

\bibitem[{Meng and Zeng(2016)}]{meng2016profit}
\bibinfo{author}{F.-L. Meng}, \bibinfo{author}{X.-J. Zeng}, \bibinfo{title}{A
  Profit Maximization Approach to Demand Response Management with Customers
  Behavior Learning in Smart Grid}, \bibinfo{journal}{IEEE Transactions on
  Smart Grid} \bibinfo{volume}{7}~(\bibinfo{number}{3}) (\bibinfo{year}{2016})
  \bibinfo{pages}{1516--1529}.

\bibitem[{Mohsenian-Rad and Leon-Garcia(2010)}]{mohsenian2010optimal}
\bibinfo{author}{A.-H. Mohsenian-Rad}, \bibinfo{author}{A.~Leon-Garcia},
  \bibinfo{title}{Optimal residential load control with price prediction in
  real-time electricity pricing environments}, \bibinfo{journal}{IEEE
  Transactions on Smart Grid} \bibinfo{volume}{1}~(\bibinfo{number}{2})
  (\bibinfo{year}{2010}) \bibinfo{pages}{120--133}.

\bibitem[{Mohsenian-Rad et~al.(2010)Mohsenian-Rad, Wong, Jatskevich, Schober,
  and Leon-Garcia}]{mohsenian2010autonomous}
\bibinfo{author}{A.-H. Mohsenian-Rad}, \bibinfo{author}{V.~W. Wong},
  \bibinfo{author}{J.~Jatskevich}, \bibinfo{author}{R.~Schober},
  \bibinfo{author}{A.~Leon-Garcia}, \bibinfo{title}{Autonomous demand-side
  management based on game-theoretic energy consumption scheduling for the
  future smart grid}, \bibinfo{journal}{IEEE Transactions on Smart Grid}
  \bibinfo{volume}{1}~(\bibinfo{number}{3}) (\bibinfo{year}{2010})
  \bibinfo{pages}{320--331}.

\bibitem[{Paterakis et~al.(2015)Paterakis, Erdinc, Bakirtzis, and
  Catal\~{a}o}]{paterakis2015optimal}
\bibinfo{author}{N.~G. Paterakis}, \bibinfo{author}{O.~Erdinc},
  \bibinfo{author}{A.~G. Bakirtzis}, \bibinfo{author}{J.~a.~P. Catal\~{a}o},
  \bibinfo{title}{Optimal household appliances scheduling under day-ahead
  pricing and load-shaping demand response strategies}, \bibinfo{journal}{IEEE
  Transactions on Industrial Informatics}
  \bibinfo{volume}{11}~(\bibinfo{number}{6}) (\bibinfo{year}{2015})
  \bibinfo{pages}{1509--1519}.

\bibitem[{Qian et~al.(2013)Qian, Zhang, Huang, and Wu}]{qian2013demand}
\bibinfo{author}{L.~P. Qian}, \bibinfo{author}{Y.~J.~A. Zhang},
  \bibinfo{author}{J.~Huang}, \bibinfo{author}{Y.~Wu}, \bibinfo{title}{Demand
  response management via real-time electricity price control in smart grids},
  \bibinfo{journal}{IEEE Journal on Selected areas in Communications}
  \bibinfo{volume}{31}~(\bibinfo{number}{7}) (\bibinfo{year}{2013})
  \bibinfo{pages}{1268--1280}.

\bibitem[{Qu et~al.(2016)Qu, Liang, Zhu, Wang, and Suganthan}]{qu2016economic}
\bibinfo{author}{B.~Qu}, \bibinfo{author}{J.~Liang}, \bibinfo{author}{Y.~Zhu},
  \bibinfo{author}{Z.~Wang}, \bibinfo{author}{P.~N. Suganthan},
  \bibinfo{title}{Economic emission dispatch problems with stochastic wind
  power using summation based multi-objective evolutionary algorithm},
  \bibinfo{journal}{Information Sciences} \bibinfo{volume}{351}
  (\bibinfo{year}{2016}) \bibinfo{pages}{48--66}.

\bibitem[{Ren et~al.(2016)Ren, Suganthan, Srikanth, and
  Amaratunga}]{ren2016random}
\bibinfo{author}{Y.~Ren}, \bibinfo{author}{P.~N. Suganthan},
  \bibinfo{author}{N.~Srikanth}, \bibinfo{author}{G.~Amaratunga},
  \bibinfo{title}{Random vector functional link network for short-term
  electricity load demand forecasting}, \bibinfo{journal}{Information Sciences}
  \bibinfo{volume}{367} (\bibinfo{year}{2016}) \bibinfo{pages}{1078--1093}.

\bibitem[{Samadi et~al.(2010)Samadi, Mohsenian-Rad, Schober, Wong, and
  Jatskevich}]{samadi2010optimal}
\bibinfo{author}{P.~Samadi}, \bibinfo{author}{A.-H. Mohsenian-Rad},
  \bibinfo{author}{R.~Schober}, \bibinfo{author}{V.~W. Wong},
  \bibinfo{author}{J.~Jatskevich}, \bibinfo{title}{Optimal real-time pricing
  algorithm based on utility maximization for smart grid}, in:
  \bibinfo{booktitle}{Smart Grid Communications (SmartGridComm), 2010 First
  IEEE International Conference on}, \bibinfo{organization}{IEEE},
  \bibinfo{pages}{415--420}, \bibinfo{year}{2010}.

\bibitem[{Siano(2014)}]{siano2014demand}
\bibinfo{author}{P.~Siano}, \bibinfo{title}{Demand response and smart grids --
  A survey}, \bibinfo{journal}{Renewable and Sustainable Energy Reviews}
  \bibinfo{volume}{30} (\bibinfo{year}{2014}) \bibinfo{pages}{461--478}.

\bibitem[{Sinha et~al.(2017)Sinha, Malo, and Deb}]{sinha2017review}
\bibinfo{author}{A.~Sinha}, \bibinfo{author}{P.~Malo},
  \bibinfo{author}{K.~Deb}, \bibinfo{title}{A Review on Bilevel Optimization:
  From Classical to Evolutionary Approaches and Applications},
  \bibinfo{journal}{IEEE Transactions on Evolutionary Computation}
  (\bibinfo{year}{2017}) \bibinfo{pages}{(in press)}.

\bibitem[{Srinivasan et~al.(2017)Srinivasan, Rajgarhia, Radhakrishnan, Sharma,
  and Khincha}]{srinivasan2017game}
\bibinfo{author}{D.~Srinivasan}, \bibinfo{author}{S.~Rajgarhia},
  \bibinfo{author}{B.~M. Radhakrishnan}, \bibinfo{author}{A.~Sharma},
  \bibinfo{author}{H.~Khincha}, \bibinfo{title}{Game-Theory based dynamic
  pricing strategies for demand side management in smart grids},
  \bibinfo{journal}{Energy} \bibinfo{volume}{126} (\bibinfo{year}{2017})
  \bibinfo{pages}{132--143}.

\bibitem[{Thimmapuram and Kim(2013)}]{thimmapuram2013consumers}
\bibinfo{author}{P.~R. Thimmapuram}, \bibinfo{author}{J.~Kim},
  \bibinfo{title}{Consumers' price elasticity of demand modeling with economic
  effects on electricity markets using an agent-based model},
  \bibinfo{journal}{IEEE Transactions on Smart Grid}
  \bibinfo{volume}{4}~(\bibinfo{number}{1}) (\bibinfo{year}{2013})
  \bibinfo{pages}{390--397}.

\bibitem[{Trivedi et~al.(2016)Trivedi, Srinivasan, Biswas, and
  Reindl}]{trivedi2016genetic}
\bibinfo{author}{A.~Trivedi}, \bibinfo{author}{D.~Srinivasan},
  \bibinfo{author}{S.~Biswas}, \bibinfo{author}{T.~Reindl}, \bibinfo{title}{A
  genetic algorithm--differential evolution based hybrid framework: Case study
  on unit commitment scheduling problem}, \bibinfo{journal}{Information
  Sciences} \bibinfo{volume}{354} (\bibinfo{year}{2016})
  \bibinfo{pages}{275--300}.

\bibitem[{Vetter et~al.(2005)Vetter, Nov\'{a}k, Wagner, Veit, M\"{o}ller,
  Besenhard, Winter, Wohlfahrt-Mehrens, Vogler, and
  Hammouche}]{vetter2005ageing}
\bibinfo{author}{J.~Vetter}, \bibinfo{author}{P.~Nov\'{a}k},
  \bibinfo{author}{M.~Wagner}, \bibinfo{author}{C.~Veit},
  \bibinfo{author}{K.-C. M\"{o}ller}, \bibinfo{author}{J.~Besenhard},
  \bibinfo{author}{M.~Winter}, \bibinfo{author}{M.~Wohlfahrt-Mehrens},
  \bibinfo{author}{C.~Vogler}, \bibinfo{author}{A.~Hammouche},
  \bibinfo{title}{Ageing mechanisms in lithium-ion batteries},
  \bibinfo{journal}{Journal of Power Sources}
  \bibinfo{volume}{147}~(\bibinfo{number}{1}) (\bibinfo{year}{2005})
  \bibinfo{pages}{269--281}.

\bibitem[{Wei et~al.(2015)Wei, Liu, and Mei}]{wei2015energy}
\bibinfo{author}{W.~Wei}, \bibinfo{author}{F.~Liu}, \bibinfo{author}{S.~Mei},
  \bibinfo{title}{Energy pricing and dispatch for smart grid retailers under
  demand response and market price uncertainty}, \bibinfo{journal}{IEEE
  Transactions on Smart Grid} \bibinfo{volume}{6}~(\bibinfo{number}{3})
  (\bibinfo{year}{2015}) \bibinfo{pages}{1364--1374}.

\bibitem[{Wu et~al.(2014)Wu, Zhou, Li, and Zhang}]{wu2014real}
\bibinfo{author}{Z.~Wu}, \bibinfo{author}{S.~Zhou}, \bibinfo{author}{J.~Li},
  \bibinfo{author}{X.-P. Zhang}, \bibinfo{title}{Real-time scheduling of
  residential appliances via conditional risk-at-value}, \bibinfo{journal}{IEEE
  Transactions on Smart Grid} \bibinfo{volume}{5}~(\bibinfo{number}{3})
  (\bibinfo{year}{2014}) \bibinfo{pages}{1282--1291}.

\bibitem[{Yu and Hong(2016{\natexlab{a}})}]{yu2016real}
\bibinfo{author}{M.~Yu}, \bibinfo{author}{S.~H. Hong}, \bibinfo{title}{A
  real-time demand-response algorithm for smart grids: A stackelberg game
  approach}, \bibinfo{journal}{IEEE Transactions on Smart Grid}
  \bibinfo{volume}{7}~(\bibinfo{number}{2})
  (\bibinfo{year}{2016}{\natexlab{a}}) \bibinfo{pages}{879--888}.

\bibitem[{Yu and Hong(2016{\natexlab{b}})}]{yu2016supply}
\bibinfo{author}{M.~Yu}, \bibinfo{author}{S.~H. Hong},
  \bibinfo{title}{Supply--demand balancing for power management in smart grid:
  A Stackelberg game approach}, \bibinfo{journal}{Applied Energy}
  \bibinfo{volume}{164} (\bibinfo{year}{2016}{\natexlab{b}})
  \bibinfo{pages}{702--710}.

\bibitem[{Yun et~al.(2008)Yun, Quan, Caixin, Shaolan, Yuming, and
  Yang}]{yun2008rbf}
\bibinfo{author}{Z.~Yun}, \bibinfo{author}{Z.~Quan},
  \bibinfo{author}{S.~Caixin}, \bibinfo{author}{L.~Shaolan},
  \bibinfo{author}{L.~Yuming}, \bibinfo{author}{S.~Yang}, \bibinfo{title}{RBF
  neural network and ANFIS-based short-term load forecasting approach in
  real-time price environment}, \bibinfo{journal}{IEEE Transactions on Power
  Systems} \bibinfo{volume}{23}~(\bibinfo{number}{3}) (\bibinfo{year}{2008})
  \bibinfo{pages}{853--858}.

\bibitem[{Zhang et~al.(2016)Zhang, Wang, Zhang, Liu, and Guo}]{zhang2016model}
\bibinfo{author}{Y.~Zhang}, \bibinfo{author}{R.~Wang},
  \bibinfo{author}{T.~Zhang}, \bibinfo{author}{Y.~Liu},
  \bibinfo{author}{B.~Guo}, \bibinfo{title}{Model predictive control-based
  operation management for a residential microgrid with considering forecast
  uncertainties and demand response strategies}, \bibinfo{journal}{IET
  Generation, Transmission \& Distribution}
  \bibinfo{volume}{10}~(\bibinfo{number}{10}) (\bibinfo{year}{2016})
  \bibinfo{pages}{2367--2378}.

\bibitem[{Zugno et~al.(2013)Zugno, Morales, Pinson, and
  Madsen}]{zugno2013bilevel}
\bibinfo{author}{M.~Zugno}, \bibinfo{author}{J.~M. Morales},
  \bibinfo{author}{P.~Pinson}, \bibinfo{author}{H.~Madsen}, \bibinfo{title}{A
  bilevel model for electricity retailers' participation in a demand response
  market environment}, \bibinfo{journal}{Energy Economics} \bibinfo{volume}{36}
  (\bibinfo{year}{2013}) \bibinfo{pages}{182--197}.

\end{thebibliography}

\end{document}